\documentclass[aps,pra,reprint,nofootinbib,floatfix,superscriptaddress]{revtex4-2}
\usepackage{mathtools,amssymb,amsthm,bm,bbm,xcolor,mathdots,stmaryrd}
\usepackage[colorlinks=true, linkcolor=blue, citecolor=magenta, urlcolor=blue]{hyperref}
\pdfoutput=1
\usepackage[T1]{fontenc}
\usepackage{lmodern}
\usepackage[utf8]{inputenc}
\usepackage{microtype}
\usepackage{fnpct}
\usepackage{dsfont}
\usepackage{orcidlink}

\theoremstyle{remark}
\newtheorem{rmk}{Remark}
\newtheorem{ex}{Example}

\newcommand{\HH}{\mathcal{H}}
\newcommand{\AB}{\mathcal{A}_{\textsc{b}}}
\newcommand{\AU}{\mathcal{A}_{\textsc{u}}}
\newcommand{\U}{\mathrm{U}}
\newcommand{\uu}{\mathfrak{u}}
\newcommand{\hh}{\mathfrak{h}}
\newcommand{\HHH}{\mathbf{H}}
\newcommand{\F}{\mathcal{F}}
\newcommand{\gH}{{g}_{\mathfrak{h}}}
\newcommand{\gFS}{{g}_{\textsc{fs}}}
\newcommand{\gWY}{{g}_{\textsc{wy}}}
\newcommand{\gHS}{{g}_{\textsc{hs}}}
\newcommand{\gG}{{g}_{\textsc{g}}}
\newcommand{\gGj}{{g}_{\textsc{g};j}}
\newcommand{\gGett}{{g}_{\textsc{g};1}}
\newcommand{\gGtva}{{g}_{\textsc{g};2}}
\newcommand{\gGm}{{g}_{\textsc{g};\bar m}}
\newcommand{\gpG}{{g}_{\times\textsc{g}}}
\newcommand{\gB}{{g}_{\textsc{b}}}
\newcommand{\gW}{{g}_{\textsc{w}}}
\newcommand{\bgW}{{\bar g}_{\textsc{w}}}
\newcommand{\gP}{{g}_{\mathbf{p}}}
\newcommand{\p}{\mathbf{p}}
\newcommand{\tr}{\operatorname{tr}}
\newcommand{\dt}{\mathrm{d}t}
\newcommand{\1}{\mathbbm{1}}
\newcommand{\bra}[1]{\langle #1|}
\newcommand{\ket}[1]{|#1\rangle}
\newcommand{\braket}[2]{\langle #1 | #2 \rangle}
\newcommand{\ketbra}[2]{|#1 \rangle\langle #2|}
\newcommand{\spec}{\operatorname{spec}}
\newcommand{\eps}{\epsilon}
\newcommand{\texp}{\mathrm{T}\!\exp}
\newcommand{\supp}{\operatorname{supp}}
\newcommand{\width}{\operatorname{width}}
\renewcommand{\P}{\mathbf{P}}
\newcommand{\wpP}{\wp_{\mathbf{p}}}
\newcommand{\bwedge}{\boldsymbol{\wedge}}
\newcommand{\tauMT}{\tau_{\textsc{mt}}}
\newcommand{\tauU}{\tau_{\textsc{u}}}
\newcommand{\tauG}{\tau_{\textsc{g}}}
\newcommand{\tauWY}{\tau_{\textsc{wy}}}
\newcommand{\tauFS}{\tau_{\textsc{fs}}}
\newcommand{\tauQSL}{\tau_{\textsc{qsl}}}
\newcommand{\tauP}{\tau_{\mathbf{p}}}
\newcommand{\dist}{\operatorname{dist}}
\newcommand{\distFS}{\operatorname{dist}_{\textsc{fs}}}
\newcommand{\distWY}{\operatorname{dist}_{\textsc{wy}}}
\newcommand{\distG}{\operatorname{dist}_{\textsc{g}}}
\newcommand{\distGj}{\operatorname{dist}_{\textsc{g};j}}
\newcommand{\distpG}{\operatorname{dist}_{\times\textsc{g}}}
\newcommand{\distB}{\operatorname{angle}_{\textsc{b}}}
\newcommand{\length}{\operatorname{length}}
\newcommand{\lengthFS}{\operatorname{length}_{\textsc{fs}}}
\newcommand{\lengthG}{\operatorname{length}_{\textsc{g}}}

\begin{document}
\title{Extensions of the Mandelstam-Tamm quantum speed limit to systems in mixed states}
\author{Niklas H{\"o}rnedal\,\orcidlink{0000-0002-2005-8694}\,}
\affiliation{Department of Physics, Stockholm University, 106 91 Stockholm, Sweden}
\affiliation{Department of Physics and Materials Science, University of Luxembourg, L-1511 Luxembourg, Luxembourg}
\author{Dan Allan\,\orcidlink{0000-0001-8437-2896}\,}
\affiliation{Department of Physics, Stockholm University, 106 91 Stockholm, Sweden}
\affiliation{Department of Physics and Materials Science, University of Luxembourg, L-1511 Luxembourg, Luxembourg}
\author{Ole S{\"o}nnerborn\,\orcidlink{0000-0002-1726-4892}\,}
\email{ole.sonnerborn@kau.se}
\affiliation{Department of Physics, Stockholm University, 106 91 Stockholm, Sweden}
\affiliation{Department of Mathematics and Computer Science, Karlstad University, 651 88 Karlstad, Sweden}

\begin{abstract}
The Mandelstam-Tamm quantum speed limit puts a bound on how fast a closed system in a pure state can evolve. In this paper, we derive several extensions of this quantum speed limit to closed systems in mixed states. We also compare the strengths of these extensions and examine their tightness. The most widely used extension of the Mandelstam-Tamm quantum speed limit originates in Uhlmann's energy dispersion estimate. We carefully analyze the underlying geometry of this estimate, an analysis that makes apparent that the Bures metric, or equivalently the quantum Fisher information, will rarely give rise to tight extensions. This observation leads us to address whether there is a tightest general extension of the Mandelstam-Tamm quantum speed limit. Using a geometric construction similar to that developed by Uhlmann, we prove that this is indeed the case. In addition, we show that tight evolutions of mixed states are typically generated by time-varying Hamiltonians, which contrasts with the case for systems in pure states.
\end{abstract}

\date{\today}
\maketitle

\section{Introduction}
A quantum speed limit is a relatively new concept in quantum mechanics that has its origins in Mandelstam and Tamm's time-energy uncertainty relation \cite{MaTa1945}.\footnote{There are several time-energy uncertainty relations in quantum theory that relate different notions of time to different notions of energy uncertainty; consult \cite{Bu2008} for more information.} According to this relation, an isolated system cannot evolve between two fully distinguishable states in a time shorter than $\pi\hbar/2$ divided by the energy uncertainty, 
\begin{equation}\label{jajjamen}
    \Delta t \geq \frac{\pi\hbar}{2\Delta E}.
\end{equation}
The bound on the right thus limits how fast an isolated quantum system can develop. Since Mandelstam and Tamm's relation can be derived from fundamental quantum mechanical principles alone, the bound is a universal inherent constraint of isolated quantum systems. Such a time-bound is called a quantum speed limit (QSL). 

Half a century after the publication of \cite{MaTa1945}, Anandan and Aharonov \cite{AnAh1990} showed that
\begin{equation}\label{hmm}
    \Delta t\geq \frac{\hbar \arccos|\braket{\psi_0}{\psi_1}|}{\Delta E}.
\end{equation}
The inverse trigonometric factor in the numerator is the Fubini-Study geodesic distance between the system's initial and final state. This distance is $\pi/2$ if these states are fully distinguishable; hence \eqref{hmm} implies \eqref{jajjamen}. The estimate \eqref{hmm} is more general than \eqref{jajjamen} because it also applies to not fully distinguishable states. Furthermore, \eqref{hmm} is valid for closed systems as well, provided that the energy uncertainty is replaced by its time average.

The lower bound in Anandan and Aharonov's estimate is called the Mandelstam-Tamm QSL, although neither Mandelstam and Tamm nor Anandan and Aharonov used this term. Since the publication of \cite{AnAh1990}, many QSLs have been derived for both open and closed systems, in pure or mixed states; see, e.g., the review articles \cite{Fr2016, DeCa2017} and the references therein. Common to all of these is that they bound the evolution time in terms of different statistical quantities such as average energy, purity, and entropy.

The most well-known and widely applied extension of the Mandelstam-Tamm QSL to systems in mixed states is due to Uhlmann \cite{Uh1992b}. In a series of papers \cite{Uh1986,Uh1989,Uh1991,Uh1992a}, Uhlmann generalized the geometric construction of Anandan and Aharonov to apply to systems in mixed states. Uhlmann then showed that the time it takes for a closed system to evolve between two mixed states is bounded from below by the fraction of $\hbar$ times the Bures angle between the states and the time average of the energy uncertainty,
\begin{equation}\label{Ulles}
	\Delta t\geq \frac{ \hbar \arccos\tr|\sqrt{\rho_0}\sqrt{\rho_1}\,|}{\Delta E}.
\end{equation}
Since mixed states are geometrically more complex than pure states, more can be said about \eqref{Ulles} than \eqref{hmm}. We will examine Uhlmann's QSL in some detail in this paper.

As an estimate of the evolution time, a QSL can be more or less tight. Many papers have been published in which QSLs are derived for different types of systems with different types of constraints. However, few of them contain a more in-depth analysis of how tight these QSLs are (\cite{GiLlMa2003,Br2003,BrHo2006,LeTo2009} are notable exceptions). To perform such an analysis is important given the potential practical significance of QSLs \cite{Ll2000,Ll2002,He2013,De2020,LaPeGrAlRoMeNeMoCaAl2021}. Mandelstam and Tamm's QSL is tight in a relatively strong sense: For every pair of pure states, there is a Hamiltonian that transforms one state to the other in a time equal to the Mandelstam-Tamm QSL. We show that this is almost never true for the Uhlmann QSL. More precisely, we show that the Uhlmann QSL is not tight under any circumstances if the rank of the initial state exceeds half the dimension of the system and that it is tight only in exceptional cases if the rank is between 2 and half of the dimension of the system.

An extension of Mandelstam and Tamm's QSL to systems in mixed states is not unique; there are many non-equivalent extensions that all reduce to the Mandelstam-Tamm QSL for pure states. Here we derive several such extensions. We also compare the strengths of these and examine their tightness. Due to its widespread use, a special section is devoted to the Uhlmann QSL. In the last section of the paper, we address whether there is a tightest general extension of the Mandelstam-Tamm QSL. The question is answered affirmatively, but some work remains to make the answer useful in practice.

The paper is organized as follows. In Section \ref{theMTQSL} we introduce terminology, review Anandan and Aharonov's derivation of the Mandelstam-Tamm QSL, and specify the type of extensions of this QSL that we will consider in this paper. In Section \ref{extfromGrass} we derive two extensions of the Mandelstam-Tamm QSL using Grassmannian geometry. Section \ref{Ulam} reviews and examines the tightness of Uhlmann's QSL. Section \ref{Wigge} discusses a QSL that is related to the Uhlmann QSL but which does not fully qualify as an extension of the Mandelstam-Tamm QSL. In Section \ref{Unitorbits} we address the question of whether there exists a tightest general extension of the Mandelstam-Tamm QSL. We prove that this is the case, and we also provide a procedure for constructing Hamiltonians that generate tight evolutions. The paper ends with a summary. 

\section{The Mandelstam-Tamm QSL}\label{theMTQSL}
In this introductory section, we review Anandan and Aharonov's derivation of the Mandelstam-Tamm QSL, and we make precise what type of extensions of this QSL we will consider in the following sections. First, however, we will introduce some terminology and notation that we will use throughout the paper.

\subsection{Terminology and notation}
All quantum systems considered in this paper will be modeled on a Hilbert space $\HH$ with a finite but otherwise unspecified dimension $N$. The states of these systems will be represented by density operators on $\HH$, that is, positive operators on $\HH$ with unit trace. We say that a state is pure if its density operator has rank $1$; otherwise we say that the state is mixed. 

The spectral decomposition of a state is the unique representation of its density operator $\rho$ as a sum $\sum_{j=1}^m p_jP_j$ where $p_1, p_2, \dots, p_m$ are the \emph{different} eigenvalues of $\rho$, tacitly assumed to be \emph{organized in descending order of magnitude}, and $P_j$ is the orthogonal projection operator onto the eigenspace of $\rho$ corresponding to $p_j$. Since the phrase ``orthogonal projection operator'' is far too long to be repeated as many times as needed, we use the term ``projector'' when referring to such an operator.

An evolving quantum state will be represented by a curve of density operators, $\rho_t$. Unless otherwise stated, we assume that $t$ ranges from an initial $t_0$ to a final $t_1$. The difference $\Delta t=t_1-t_0$ is the evolution time. For simplicity, we will often write $\rho_0$ for the initial state $\rho_{t_0}$ and $\rho_1$ for the final state $\rho_{t_1}$. Furthermore, we will write $\sum_{j=1}^m p_jP_{j;t}$ for the spectral decomposition of $\rho_t$. Here, the eigenvalues will not depend on $t$ since we will only consider closed systems, that is, systems such that $\rho_t$ satisfies a von Neumann equation with a Hamiltonian that may depend on time. For such a system, only the eigenspace projectors change. 

Time averages of functions defined on the evolution time interval are common in this paper. For this reason, we introduce a special notation for the time average: If $f_t$ is defined on the evolution time interval, then
\begin{equation}
    \langle f_t\rangle 
    = \frac{1}{\Delta t}\int_{t_0}^{t_1}\dt f_t.
\end{equation}
An example of such a function is the energy uncertainty along an evolution curve:
\begin{equation}
    \Delta(H_t,\rho_t)=\sqrt{\tr(H_t^2\rho_t) - \tr(H_t\rho_t)^2}.
\end{equation}
For the time average of this function, we introduce the special notation
\begin{equation}
    \Delta E
    =\langle\Delta(H_t,\rho_t)\rangle.
\end{equation}

We will introduce several smooth manifolds of quantum states. A tangent vector at a state $\rho$ in such a manifold will typically be denoted by $\dot\rho$, and when we need to distinguish between different tangent vectors at $\rho$ we will index them with Latin letters. The manifolds will be equipped with Riemannian metrics. Following tradition, a Riemannian metric will be denoted by $g$ and the inner product of two tangent vectors $\dot\rho_a$ and $\dot\rho_b$ by $g(\dot\rho_a,\dot\rho_b)$. We write $\|\dot\rho\|_g$ for the size of $\dot\rho$, that is, $\|\dot\rho\|_g=\sqrt{g(\dot\rho,\dot\rho)}$. By definition, the length of a curve of states $\rho_t$ is the integral of its speed $\|\dot\rho_t\|_g$, or equivalently the evolution time multiplied with the average speed, 
\begin{equation}
    \length_g[\rho_t]
    = \int_{t_0}^{t_1}\dt\,\|\dot\rho_t\|_g
    = \Delta t \langle \|\dot\rho_t\|_g\rangle.
\end{equation}
The geodesic distance between two states is the infimum of all the lengths of smooth curves that connect them,
\begin{equation}
    \dist_g(\rho_0,\rho_1)
    = \inf\{\Delta t \langle \|\dot\rho_t\|_g\rangle : 
    \rho_{t_0}=\rho_0,\, \rho_{t_1}=\rho_1\}. 
\end{equation}

Finally, we will use the convention that the absolute value of an operator is defined as $|A|=\sqrt{AA^\dagger}$. Moreover, all quantities are expressed in units such that $\hbar$ assumes the value $1$.

\subsection{The Mandelstam-Tamm QSL in a nutshell}
Pure quantum states are represented by projectors with $1$-dimensional support. The projective Hilbert space $P(\HH)$ is the manifold of all such projectors on $\HH$. This manifold is compact and admits an essentially unique unitarily invariant Riemannian metric 
\begin{equation}\label{FSmetric}
	\gFS(\dot\rho_a,\dot\rho_b) = \frac{1}{2}\tr(\dot\rho_a\dot\rho_b)
\end{equation}
called the Fubini-Study metric.

Consider a closed system modeled on $\HH$ with Hamiltonian $H_t$, and assume that the system's state is pure. When the system evolves, the state's density operator traces a curve $\rho_t$ in $P(\HH)$. The square of the Fubini-Study speed of $\rho_t$ equals the energy variance:
\begin{equation}\label{FSvariance}
\begin{split}
	\|\dot\rho_t\|_{\gFS}^2
	= \frac{1}{2}\tr\big((-i[H_t,\rho_t])^2\big)
	= \Delta^2\left(H_t,\rho_t\right).
\end{split}
\end{equation}
Hence, the length of $\rho_t$ equals the evolution time multiplied by the average energy uncertainty,
\begin{equation}\label{FSlangd}
\begin{split}
	\lengthFS[\rho_t]
	= \Delta t\Delta E.
\end{split}
\end{equation}
The length of $\rho_t$ is greater than or equal to the geodesic distance between the initial state $\rho_0=\rho_{t_0}$ and the final state $\rho_1=\rho_{t_1}$. This distance is
\begin{equation}\label{FSdistance}
	\distFS(\rho_0,\rho_1)
	= \arccos\sqrt{\tr(\rho_0\rho_1)};
\end{equation}
see Section \ref{hopf}. Thus, the evolution time is bounded from below by the geodesic distance divided by the average energy uncertainty,
\begin{equation}\label{MTQSL}
	\Delta t 
	\geq \frac{\arccos\sqrt{\tr(\rho_0\rho_1)}}{\Delta E}.
\end{equation}
The bound is the Mandelstam-Tamm QSL, which we hereafter denote by $\tauMT$.

\subsection{Extensions of the Mandelstam-Tamm QSL}
In this paper we derive and compare extensions of the Mandelstam-Tamm QSL to closed systems in mixed states. The extensions will all be of the form 
\begin{equation}\label{genqsl}
	\tauQSL = \frac{d(\rho_0,\rho_1)}{\Delta E},
\end{equation} 
where $\rho_0$ and $\rho_1$ represent the initial and the final state of the system, respectively, and $\Delta E$ is the average energy uncertainty along the evolution curve. The function $d$ does not have to be a distance function. However, we require that $d$ only takes pairs of density operators as an argument and that $d(\rho_0,\rho_1)$ is the Fubini-Study geodesic distance between $\rho_0$ and $\rho_1$ if these represent pure states.

We will also discuss the tightness of the QSLs we derive. We say that a QSL is tight if there is a Hamiltonian that transforms the initial state to the final state in a time that coincides with the QSL. That the Mandelstam-Tamm QSL is tight follows from \eqref{FSlangd} and that every geodesic of the Fubini-Study metric is generated by a Hamiltonian.

\section{Extensions of the Mandelstam-Tamm QSL from geometries on Grassmannians}\label{extfromGrass}
When a quantum state evolves unitarily, its eigenvalue spectrum is kept intact, and its eigenspace projectors rotate without changing their rank. Each rotating projector traces a curve in a Grassmann manifold. Grassmann manifolds are the equivalents of the projective Hilbert space for higher-dimensional projectors. In this section we discuss some geometric properties of dynamical systems in Grassmann manifolds. We then use these properties to derive two extensions of the Mandelstam-Tamm QSL.

\subsection{Speed and distance in Grassmannians}
The Grassmann manifold $G(n,\HH)$ is the manifold of projectors on $\HH$ having rank $n$. We equip $G(n,\HH)$ with the Riemannian metric
\begin{equation}\label{Grassmann_metric}
	\gG(\dot P_a,\dot P_b)
	= \frac{1}{2}\tr(\dot P_a\dot P_b),
\end{equation}
henceforth referred to as the Grassmann metric. The square of the speed of a curve $P_t$ in $G(n,\HH)$ generated by a Hamiltonian $H_t$ is
\begin{equation}\label{Gspeed}
\begin{split}
	\|\dot P_t\|^2_{\gG} 
	&= \frac{1}{2}\tr\big((-i[H_t,P_t])^2\big) \\
	&= \tr(H_t^2P_t)-\tr(H_tP_tH_tP_t).    
\end{split}
\end{equation}
For reasons that will be explained in the next section, we write $I(H_t,P_t)$ for the speed of $P_t$ squared. 

Assume that $t$ ranges from $t_0$ to $t_1$, and let $P_0=P_{t_0}$ and $P_1=P_{t_1}$. The length of $P_t$ is equal to the duration of the evolution, $\Delta t=t_1 - t_0$, times the average value of the square root of $I(H_t,P_t)$,
\begin{equation}
	\lengthG[P_t]
	= \Delta t \big\langle\! \sqrt{I(H_t,P_t)} \,\big\rangle.
\end{equation}
The length is greater than or equal to the geodesic distance between the initial and final projector. In Appendix \ref{grassmann_distance} we show that this distance is
\begin{equation}\label{distanceG}
	\distG(P_0,P_1)
	=\sqrt{\tr\arccos^2|P_0P_1|-\frac{\pi^2}{4}(N-n)}.
\end{equation}
Edelman et al.\ \cite{EdArSm1998} derived an equivalent formula for the geodesic distance using Stiefel frame bundles; see equation \eqref{diastnceGII} and Remark \ref{Stiefels} in Section \ref{hopf}. 

\subsection{Skew information}
Wigner and Yanase \cite{WiYa1963} proposed the skew information 
\begin{equation}
	I(A,\rho)
	= \tr (A^2\rho)-\tr(A\sqrt{\rho}A\sqrt{\rho})
\end{equation}
as a measure of the amount of information that a state $\rho$ contains with respect to an observable $A$. Interestingly, if the observable is the Hamiltonian of a closed system in a faithful state, the square root of twice the skew information equals the speed with which the state evolves according to a particular metric \cite{GiIs2003}. In Section \ref{Wigge} we show that under certain assumptions on the spectral width, this observation leads to a QSL similar to that of Mandelstam and Tamm. 

Luo \cite{Lu2003} pointed out that the skew information of a state relative to an observable never exceeds the variance of the observable, $I(A,\rho) \leq \Delta^2(A,\rho)$. The two quantities agree for pure states, but for mixed states the skew information is strictly smaller than the variance. Here we refine this observation somewhat.

Define the skew information of a projector $P$ relative to an observable $A$ as 
\begin{equation}
	I(A,P) = \tr(A^2P)-\tr(APAP).
\end{equation}
Furthermore, for a density operator $\rho$ with spectral decomposition $\sum_{j=1}^m p_j P_j$ define $J(A,\rho)$ as
\begin{equation}\label{jay}
	J(A,\rho) = \sum_{j=1}^m p_j I(A,P_j).
\end{equation}
This quantity is greater than the skew information of $\rho$ but less than the variance of $A$,
\begin{equation}\label{duo}
	I(A,\rho)
	\leq J(A,\rho)
	\leq \Delta^2(A,\rho).
\end{equation} 
These inequalities follow from
\begin{equation}\label{A1}
	\tr(A\rho)^2 
	\leq \sum_{j=1}^m p_j \tr(A P_j A P_j) 
	\leq \tr(A\sqrt{\rho} A\sqrt{\rho}).
\end{equation}
To prove this latter pair of inequalities, let $n_j$ be the rank of $P_j$ and let $\ket{j;1}, \ket{j;2},\dots,\ket{j;n_j}$ be an orthonormal set of vectors spanning the support of $P_j$. Then
\begin{equation}
\begin{split}
	\tr(A\rho)^2 
	&= \Big( \sum_{j=1}^m \sum_{k=1}^{n_j} p_j \bra{j;k} A \ket{j;k} \Big)^2 \\
	&\leq \sum_{j=1}^m\sum_{k=1}^{n_j} p_j \bra{j;k} A \ket{j;k}^2 \\
	&\leq \sum_{j=1}^m\sum_{k,l=1}^{n_j} p_j |\bra{j;k} A \ket{j;l}|^2 \\
	&\leq \sum_{i,j=1}^m\sum_{k,l=1}^{n_j} \sqrt{p_ip_j}\, |\bra{i;k} A \ket{j;l}|^2.
\end{split}
\end{equation}
The first inequality follows from the convexity of the squaring function, the triple sum in the third line equals $\sum_{j=1}^m p_j \tr(A P_j A P_j)$, and the quadruple sum in the last line equals $\tr(A\sqrt{\rho} A\sqrt{\rho})$.

\subsection{A quantum speed limit}\label{frost}
Let $\rho_t$ be the state of an evolving closed system with Hamiltonian $H_t$. Write $\sum_{j=1}^mp_jP_{j;t}$ for the spectral decomposition of $\rho_t$, and let $\bar m=m-1$ if $p_m=0$ and $\bar m=m$ if $p_m\ne 0$.\footnote{Recall that the $p_j$s denote the different eigenvalues of $\rho_t$ and that these are organized in descending order of magnitude.}  The eigenspace projectors corresponding to the nonzero eigenvalues collectively form a curve 
\begin{equation}\label{curve}
	\P_t=(P_{1;t},P_{2;t},\dots,P_{\bar m;t})
\end{equation}
in the $\bar m$-fold Cartesian product $\prod_{j=1}^{\bar m} G(n_j,\HH)$, where $n_j$ is the rank of $P_{j;t}$. The tangent bundle of the product splits as a direct sum of the tangent bundles of the factors, and the velocity field of $\P_t$ is the sum of the individual projectors' velocities, 
\begin{equation}
	\dot\P_t 
	= \dot P_{1;t}\oplus \dot P_{2;t}\oplus\dots\oplus \dot P_{\bar m;t}.
\end{equation}
Let $\gGj$ be the Grassmann metric on $G(n_j,\HH)$, and equip $\prod_{j=1}^{\bar m} G(n_j,\HH)$ with the weighted product metric 
\begin{equation}\label{Grassmannprodmetric}
	\gpG
	= p_1\gGett \oplus p_2\gGtva \oplus \dots \oplus p_{\bar m}\gGm.
\end{equation}
By equation \eqref{Gspeed} and the right estimate in \eqref{duo}, 
\begin{equation}\label{tjutva}
\begin{split}
   \|\dot\P_t\|^2_{\gpG}
   &= 	\sum_{j=1}^{\bar m}p_j\|\dot P_{j;t}\|^2_{\gGj} \\ 
   &= J(H_t,\rho_t) \\
   &\leq \Delta^2(H_t,\rho_t).
\end{split}
\end{equation}
Let $\distpG$ be the geodesic distance function associated with $\gpG$. Equation  \eqref{tjutva} implies that
\begin{equation}\label{Grassmanndispest}
	\distpG(\P_{t_0},\P_{t_1})\leq \Delta t\Delta E.
\end{equation}
Write $P_{j;0}$ and $P_{j;1}$ for $P_{j;t_0}$ and $P_{j;t_1}$, respectively, and let $\distGj(P_{j;0},P_{j;1})$ be the geodesic distance between $P_{j;0}$ and $P_{j;1}$ relative to the metric $\gGj$. Since the Grassmann manifolds are complete, being compact, the geodesic distance in $\prod_{j=1}^{\bar m} G(n_j,\HH)$ satisfies the Pythagorean identity
\begin{equation}\label{Pythsats}
	\distpG^2(\P_{t_0},\P_{t_1}) 
	= \sum_{j=1}^{\bar m} p_j\distGj^2(P_{j;0},P_{j;1});
\end{equation}
see \cite{Sa1996}. Equations \eqref{distanceG}, \eqref{Grassmanndispest}, and \eqref{Pythsats} yield
\begin{equation}\label{Gqsl}
	\Delta t \geq \frac{1}{\Delta E}
	\sqrt{\sum_{j=1}^{m} p_j \Big(\tr\arccos^2|P_{j;0}P_{j;1}|-\frac{\pi^2}{4}(N-n_j)\Big)}.
\end{equation}
The expression on the right is our first QSL, which we call the Grassmann QSL and denote by $\tauG$. For systems in pure states, $\tauG=\tauMT$. Thus, $\tauG$ extends the Mandelstam-Tamm QSL. Notice that we can let the upper limit of the sum be $m$ also in those cases where $\bar m = m-1$.

\begin{ex}\label{ex:distinguishable}
Two mixed states are fully distinguishable if and only if they have orthogonal supports. According to \eqref{Gqsl}, the time it takes to unitarily transform a mixed state into a fully distinguishable state is bounded from below by $\pi/2\Delta E$. This bound is tight.
\end{ex}

\begin{ex}\label{ex:non-degenerate}
Assume that $\rho_0$ and $\rho_1$ are isospectral and nondegenerate. Let 
\begin{align}
	\rho_0 &= \sum_{j=1}^N p_j\ketbra{u_j}{u_j},\\
	\rho_1 &= \sum_{j=1}^N p_j\ketbra{v_j}{v_j},
\end{align} 
be their spectral decompositions. Then
\begin{equation}\label{tusk}
	\tauG
	= \frac{1}{\Delta E} \sqrt{\sum_{j=1}^N p_j \arccos^2|\braket{u_j}{v_j}|}.
\end{equation}
\end{ex}

\begin{ex}\label{ex:non-degenerate-comm}
Suppose that $\rho_0$ and $\rho_1$ in Example \ref{ex:non-degenerate} commute. Then there is a permutation $\sigma$ of $\{1,2,\dots,N\}$ such that $\ketbra{v_j}{v_j}=\ketbra{u_{\sigma(j)}}{u_{\sigma(j)}}$. Consequently,
\begin{equation}
	|\braket{u_j}{v_j}|
	= \begin{cases}
		1 & \text{ if $\sigma(j) = j$}, \\
		0 & \text{ if $\sigma(j)\ne j$}.		
	\end{cases}
\end{equation}
In this case, the Grassmann QSL reduces to 
\begin{equation}\label{esti}
	\tauG = \frac{\pi}{2\Delta E}\sqrt{\sum_{j\ne\sigma(j)} p_j}.
\end{equation}
\end{ex}

The following example shows that the QSL in equation \eqref{esti} is tight if $\sigma$ is an involution, that is, a permutation whose square is the trivial permutation.

\begin{ex}\label{ex:two-cycles}
Every permutation of a finite set decomposes uniquely into a product of disjoint cycles; see \cite{La1993}. A permutation is an involution if and only if the permutation has cycles of length at most $2$.

Let $\rho_0$, $\rho_1$, and $\sigma$ be as in Example \ref{ex:non-degenerate-comm}, and assume that $\sigma$ is an involution. Then there is a Hamiltonian which transforms $\rho_0$ into $\rho_1$ in the time $\Delta t=\tauG$ and, hence, $\tauG$ in equation \eqref{esti} is tight. For example, the Hamiltonian 
\begin{equation}
	H = i\!\!\sum_{j\ne\sigma(j)}\!\!\big(\ketbra{u_{\sigma(j)}}{u_j} - \ketbra{u_j}{u_{\sigma(j)}}\big)
\end{equation}
transforms $\rho_0$ into $\rho_1$ in the time $\Delta t=\pi/2$. The uncertainty of $H$ is conserved, the expectation value of $H$ vanishes, and the second moment of $H$ is 
\begin{equation}
	\tr(H^2\rho_0)=\sum_{j=1}^N p_j \bra{u_j} H^2 \ket{u_j}
	= \sum_{j\ne\sigma(j)} p_j.
\end{equation}
Thus, $\Delta t$ equals $\tauG$ given by \eqref{esti}.
\end{ex}

The Grassmann distance between two projectors is the same as the Grassmann distance between their complementary projectors,
\begin{equation}\label{lika}
    \distG(P_0,P_1)=\distG(\1-P_0,\1-P_1).
\end{equation}
This follows immediately from the fact that if $P_t$ is a curve of projectors of rank $n$, then $\1-P_t$ is a curve of projectors of rank $N-n$ having the same speed, and hence the same length, as $P_t$. The next example is a consequence of \eqref{lika}.

\begin{ex}\label{two}
If the spectrum of the initial state consist of only two different eigenvalues, the Grassmann QSL is tight. To see this let
\begin{align}
    \rho_0 &= p_1P_{1;0} + p_2P_{2;0}, \\
    \rho_1 &= p_1P_{1;1} + p_2P_{2;1},
\end{align}
be the spectral decompositions of the initial and the final state, and let $H$ be a time-independent Hamiltonian which is parallel at $P_{1;0}$ and which transforms $P_{1;0}$ to $P_{1;1}$ along a shortest geodesic in the time $\Delta t=t_1-t_0=1$; see Appendix \ref{grassmann_distance}. Then, by equations \eqref{Pythsats}, \eqref{lika}, and \eqref{fyra},
\begin{equation}
\begin{split}
   \distpG^2(\P_{t_0},\P_{t_1}) 
    &= (p_1+p_2)\distG^2(P_{1;0},P_{1;1}) \\
    &= (p_1+p_2)\tr(H^2P_{1;0}).
\end{split}
\end{equation}
Since $H$ is parallel at $P_{1;0}$, the expectation value of $H$ vanishes. A second application of equation \eqref{fyra} yields 
\begin{equation}
\begin{split}
    \Delta^2E
    &= p_1 \tr(H^2P_{1;0}) + p_2 \tr(H^2P_{2;0}) \\
    &= (p_1-p_2)\tr(H^2P_{1;0}) + p_2 \tr(H^2) \\
    &= (p_1+p_2)\tr(H^2P_{1;0}).
\end{split}
\end{equation}
This shows that the Grassmann QSL is tight.
\end{ex}

The last example in this section shows that if the spectrum of the initial state consists of three or more eigenvalues, the Grassmann QSL need not be tight.

\begin{ex}\label{Grassmann_not_tight}
Assume that $N=3$ and let $\rho_0$, $\rho_1$, and $\sigma$ be as in Example \ref{ex:non-degenerate-comm}, with $\sigma$ being the permutation $(1,2,3)$:
\begin{align}
	\rho_0=p_1\ketbra{u_1}{u_1}+p_2\ketbra{u_2}{u_2}+p_3\ketbra{u_3}{u_3}, \\
	\rho_1=p_1\ketbra{u_2}{u_2}+p_2\ketbra{u_3}{u_3}+p_3\ketbra{u_1}{u_1}.
\end{align}
Contrary to what we are going to show, assume that the Grassmann QSL is tight, and let $H_t$ be a Hamiltonian that transforms $\rho_0$ into $\rho_1$ in the time $\Delta t=t_1-t_0=\tauG$. Let $\rho_t$ represent the evolving state, write $\sum_{j=1}^3 p_jP_{j;t}$ for the spectral decomposition of $\rho_t$, and without loss of generality assume that $H_t$ is such that $P_{j;t} H_t P_{j;t}=0$ for every $j$, that is, assume that $H_t$ is parallel transporting in the sense of Section \ref{tightest}. Since the Grassmann QSL is saturated, the curve of eigenspace projectors $\P_t$ is a shortest curve between $\P_{t_0}$ and $\P_{t_1}$ in $\prod_{j=1}^3P(\HH)$. Consequently, each $P_{j;t}$ is a shortest curve between $\ketbra{u_j}{u_j}$ and $\ketbra{u_{\sigma(j)}}{u_{\sigma(j)}}$ in $P(\HH)$; see \cite{Sa1996}.

Write $P_{j;t}=\ketbra{u_{j;t}}{u_{j;t}}$, where $\ket{\dot{u}_{j;t}}=-iH_t\ket{u_{j;t}}$ and $\ket{u_{j;t_0}}=\ket{u_j}$. As is shown in \cite{Br2003}, each $\ket{u_{j;t}}$ traces a curve in the linear span of $\ket{u_{j}}$ and $\ket{u_{\sigma(j)}}$,
\begin{equation}
	\ket{u_{j;t}} = \alpha_{j;t}\ket{u_{j}} + \beta_{j;t}\ket{u_{\sigma(j)}}.
\end{equation}
It follows that the field of velocity vectors $\ket{\dot{u}_{j;t}}$ is everywhere perpendicular to $\ket{{u}_{\sigma^2(j)}}$, implying that 
\begin{equation}
	\bra{{u}_{\sigma^2(j)}} H_t \ket{u_{j;t}} = i \braket{{u}_{\sigma^2(j)}}{\dot{u}_{j;t}} = 0.
\end{equation}
We thus have that
\begin{align}
	\alpha_{1;t} \bra{{u}_{3}}H_t\ket{u_{1}} + \beta_{1;t} \bra{{u}_{3}}H_t\ket{u_{2}}=0, \label{t}\\
	\alpha_{2;t} \bra{{u}_{1}}H_t\ket{u_{2}} + \beta_{2;t} \bra{{u}_{1}}H_t\ket{u_{3}}=0, \label{tt}\\
	\alpha_{3;t} \bra{{u}_{2}}H_t\ket{u_{3}} + \beta_{3;t} \bra{{u}_{2}}H_t\ket{u_{1}}=0. \label{ttt}
\end{align}
Since $\Delta t=\tauG$, the Hamiltonian $H_t$ must be nonzero for $t$ close to but greater than the initial time $t_0$. Furthermore, for such $t$, each $\alpha_{j;t}$ is close to $1$ and each $\beta_{j;t}$ is close to $0$. Now, if $\bra{{u}_{\sigma^2(j)}} H_t \ket{u_{j}}\ne 0$ for all $j$ and $t$ close to $t_0$, equations \eqref{t}-\eqref{ttt} imply that 
\begin{align}
	|\bra{{u}_{3}}H_t\ket{u_{1}}| < |\bra{{u}_{3}}H_t\ket{u_{2}}|, \label{s}\\
	|\bra{{u}_{1}}H_t\ket{u_{2}}| < |\bra{{u}_{1}}H_t\ket{u_{3}}|, \label{ss}\\
	|\bra{{u}_{2}}H_t\ket{u_{3}}| < |\bra{{u}_{2}}H_t\ket{u_{1}}|. \label{sss}
\end{align}
But these inequalities are contradictory. Consequently, $\bra{{u}_{\sigma^2(j)}} H_t \ket{u_{j}}=0$ for some $j$.
Then $\bra{{u}_{\sigma^2(j)}} H_t \ket{u_{j}}=0$ for all $j$ according to \eqref{t}-\eqref{ttt}. 
Furthermore, since $H_t$ is parallel transporting, 
\begin{equation}
	|\alpha_{j;t}|^2 \bra{{u}_{j}}H_t\ket{u_{j}} + |\beta_{j;t}|^2 \bra{{u}_{\sigma(j)}}H_t\ket{u_{\sigma(j)}}=0.\label{ruge}
\end{equation}
From this and $H_t$ being nonzero for $t$ near $t_0$ follow that
\begin{align}
	|\bra{{u}_{1}}H_t\ket{u_{1}}| < |\bra{{u}_{2}}H_t\ket{u_{2}}|, \label{u}\\
	|\bra{{u}_{2}}H_t\ket{u_{2}}| < |\bra{{u}_{3}}H_t\ket{u_{3}}|, \label{uu}\\
	|\bra{{u}_{3}}H_t\ket{u_{3}}| < |\bra{{u}_{1}}H_t\ket{u_{1}}|. \label{uuu}
\end{align}
These inequalities are contradictory. We conclude that the Grassmann QSL is not tight.
\end{ex}

\subsection{Another quantum speed limit}
Another extension of the Mandelstam-Tamm QSL can be derived using the Pl{\"u}cker embedding. This QSL is weaker than the Grassmann QSL but has the advantage of often being easier to calculate.

A frame for a projector of rank $n$ is a row matrix of $n$ pairwise orthogonal unit vectors spanning the projector's support. If $F$ is a frame for $P$, then $P=FF^\dagger$. And if $F_0$ and $F_1$ are frames for $P_0$ and $P_1$, the Grassmann geodesic distance between $P_0$ and $P_1$ can be written as
\begin{equation}\label{diastnceGII}
	\distG(P_0,P_1)
	= \sqrt{\tr\arccos^2|F_0^\dagger F_1|}.
\end{equation}

The Pl{\"u}cker embedding is an embedding of the Grassmannian $G(n,\HH)$ in the projective space over the $n$-fold alternating product $\bwedge^n\HH$; see \cite{GrHa1994}. The Pl{\"u}cker embedding identifies a projector $P$ with $\ketbra{F}{F}$, where $F$ is any frame for $P$ and $\ket{F}$ is the alternating product of the vectors in $F$. The alternating product $\bwedge^n\HH$ is a Hilbert space with the Hermitian inner product specified by the requirement that for any two frames, $\braket{F_0}{F_1}=\det F_0^\dagger F_1$. We equip the projective space $P(\bwedge^n\HH)$ with the Fubini-Study metric. In Appendix \ref{plucker_is_isometry} we show that the Pl{\"u}cker embedding is an isometry. From this follows that the Grassmann distance between any two projectors is greater than the corresponding Fubini-Study distance,
\begin{equation}\label{GgeqFS}
	 \distG(P_0,P_1) \geq  \distFS(P_0,P_1).
\end{equation}
According to equation \eqref{FSdistance}, the Fubini-Study distance between $P_0$ and $P_1$ is 
\begin{equation}\label{land}
	 \distFS(P_0,P_1) 
	 = \arccos |\det F_0^\dagger F_1|.
\end{equation}

Consider the curve in \eqref{curve} formed by the eigenspace projectors of $\rho_t$ corresponding to the nonzero eigenvalues. Let $F_{j;0}$ and $F_{j;1}$ be any frames for $P_{j;0}=P_{j;t_0}$ and $P_{j;1}=P_{j;t_1}$, respectively. By \eqref{Grassmanndispest}, \eqref{Pythsats}, \eqref{GgeqFS}, and \eqref{land},
\begin{equation}
	\Delta t
	\geq \frac{1}{\Delta E} \sqrt{\sum_{j=1}^{m} p_j \arccos^2 |\det F_{j;0}^\dagger F_{j;1}|}.
\end{equation}
The expression on the right is our second QSL, denoted by $\tauFS$. For pure states it coincides with the Mandelstam-Tamm QSL. Since the Grassmann distance is greater than the Fubini-Study distance, $\tauFS$ is weaker than $\tauG$. In general, however, $\tauFS$ is easier to calculate.

\begin{ex}\label{G vs FS}
For systems in nondegenerate states, $\tauG$ and $\tauFS$ agree; cf.\ Example \ref{ex:non-degenerate}. However, this is not the case in general. To see this let $\rho_0$ and $\rho_1$ be isospectral states with spectral decompositions
\begin{align}
	\rho_0 &= \sum_{j=1}^{m} p_jP_{j;0},\\
	\rho_1 &= \sum_{j=1}^{m} p_jP_{j;1}.
\end{align}
For each $j$ let $F_{j;0}$ and $F_{j;1}$ be frames for $P_{j;0}$ and $P_{j;1}$, respectively, and let $s_{j1},s_{j2},\dots,s_{jn_j}$ be the singular values of $F_{j;0}^\dagger F_{j;1}$. Then
\begin{align}
	\tauG
	&= \frac{1}{\Delta E}\sqrt{ \sum_{j=1}^m \sum_{k=1}^{n_j} p_j \arccos^2 s_{jk}}, \\
	\tauFS
	&= \frac{1}{\Delta E}\sqrt{ \sum_{j=1}^m p_j \arccos^2(s_{j1}s_{j2}\cdots s_{jn_j})}.
\end{align}
Since $\arccos^2x+\arccos^2y\geq \arccos^2(xy)$ for $0\leq x,y\leq 1$, with strict inequality unless $x=1$ or $y=1$, 
we have that
\begin{equation}
	\sum_{k=1}^{n_j} \arccos^2 s_{jk} 
	\geq \arccos^2(s_{j1}s_{j2}\cdots s_{jn_j})
\end{equation}
with strict inequality unless all but possibly one singular value equals $1$. The singular values that equal $1$ correspond to orthogonal directions in the intersection of the supports of $P_{j;0}$ and $P_{j;1}$. Thus, if for some nonzero $p_j$ the intersection of the supports of $P_{j;0}$ and $P_{j;1}$ has a dimension less than $n_j-1$, then $\tauG>\tauFS$.
\end{ex}

\section{The Uhlmann QSL}\label{Ulam}
When we derived the formula \eqref{distanceG} for the Grassmann geodesic distance, see Appendix \ref{grassmann_distance}, we embedded the Grassmannian in the manifold of Hermitian operators and used techniques available in subRiemannian geometry. Anandan and Aharonov \cite{AnAh1990}, on the other hand, did not consider the projective Hilbert space to be a submanifold of the Hermitian operators when deriving the Mandelstam-Tamm QSL. Instead they defined the Fubini-Study metric as a projection of a spherical metric via the Hopf bundle and employed techniques from the geometry of principal fiber bundles. Inspired by  \cite{AnAh1990}, Uhlmann \cite{Uh1986,Uh1989,Uh1991,Uh1992a} generalized the Hopf bundle to a fiber bundle over the manifold of faithful states. Uhlmann \cite{Uh1992b} then derived one of the most widely used extensions of the Mandelstam-Tamm QSL.

In this section we first repeat the definition of the generalized Hopf bundle. We then take a closer look at Uhlmann's generalization of the Hopf bundle, and we derive Uhlmann's QSL in a slightly different way than Uhlmann did. Unlike Uhlmann's, our derivation does not rely on the initial state being faithful.\footnote{A state is faithful if it is represented by a density operator whose rank equals the dimension of the Hilbert space.}\footnote{Uhlmann claimed, without a proof, that the general case follows from the faithful case by continuity. If true, this is a highly nontrivial fact since, as we will show, the Uhlmann energy dispersion estimate is never tight for systems in faithful states but can be tight for systems in states of small enough rank.} We finish with a discussion on the tightness of the Uhlmann QSL.

\subsection{The Hopf bundle}\label{hopf}
Let $W(1,\HH)$ be the sphere of unit vectors in $\HH$ equipped with the metric induced from the real part of the Hermitian product on $\HH$. The projection $\wp_1(\ket{\psi})=\ketbra{\psi}{\psi}$ from $W(1,\HH)$ onto the projective Hilbert space $P(\HH)$ is a principal fiber bundle called the Hopf bundle. The symmetry group is $\U(1)$, whose action on the fibers of $\wp_1$ is $e^{i\theta}\cdot\ket{\psi} = \ket{\psi}e^{i\theta}$.

The vertical bundle of $\wp_1$ is the kernel bundle of the differential of $\wp_1$. We take the horizontal bundle to be the orthogonal complement of the vertical bundle. The horizontal bundle is the kernel bundle of the $\uu(1)$-valued Berry connection $\AB\ket{\dot{\psi}}=\braket{\psi}{\dot{\psi}}$.

Since the symmetry group of $\wp_1$ acts by isometries, we can project the metric on $W(1,\HH)$ to a metric $g$ on $P(\HH)$. This metric is the Fubini-Study metric \eqref{FSmetric}. To see this, let $\dot\rho$ be any tangent vector at $\rho$ in $P(\HH)$, let $\ket{\psi}$ be any vector in the fiber over $\rho$, and let $\ket{\dot{\psi}}$ be a lift of $\dot\rho$ to $\ket{\psi}$. Then $\ket{\dot{\psi}^h}=\ket{\dot{\psi}}-\ket{\psi}\braket{\psi}{\dot{\psi}}$ is a horizontal lift of $\dot\rho$, and
\begin{equation}\label{snok}
	g(\dot\rho,\dot\rho) 
	= \braket{ \dot{\psi}^h }{ \dot{\psi}^h }
	= \braket{ \dot{\psi} }{ \dot{\psi} } - \braket{ \psi }{ \dot{\psi} } \braket{ \dot{\psi} }{ \psi }.
\end{equation}
Using that $\dot\rho = \ketbra{ \dot{\psi} }{ \psi } + \ketbra{ \psi }{ \dot{\psi} }$ and that $\braket{\psi}{\dot{\psi}}$ is imaginary, the right-hand side of equation \eqref{snok} is easily identified as $\tr(\dot\rho^2)/2$; thus, $g=\gFS$. If $\dot\rho=-i[H,\rho]$, and we choose $\ket{ \dot{\psi} }=-iH\ket{\psi}$, we recover the key observation \eqref{FSvariance}. 

The projection $\wp_1$ preserves the length of horizontal curves, and a curve in $P(\HH)$ is a geodesic if and only if its horizontal lifts are geodesics in $W(1,\HH)$; see \cite{Mo2006}. By definition, the Fubini-Study geodesic distance between two pure states equals the length of a shortest curve connecting the two states. This distance equals the length of a shortest curve in $W(1,\HH)$ connecting the fibers over the two states. Such a curve is a horizontal geodesic, possibly after a reparameterization \cite{Mo2006}. Spherical geometry tells us that geodesics in $W(1,\HH)$ are great arcs and that the length of the great arc in $W(1,\HH)$ connecting $\ket{\psi_0}$ and $\ket{\psi_1}$ is $\arccos\Re\braket{\psi_0}{\psi_1}$, with $\Re\braket{\psi_0}{\psi_1}$ being the real part of $\braket{\psi_0}{\psi_1}$. If $\rho_0=\ketbra{\psi_0}{\psi_0}$ and $\rho_1=\ketbra{\psi_1}{\psi_1}$, we thus have that
\begin{equation}\label{calc}
\begin{split}
	\distFS(\rho_0,\rho_1)
	&=\min_{\theta\in\mathds{R}} \arccos\Re\bra{\psi_0}e^{i\theta}\ket{\psi_1} \\
	&= \arccos (\max_{\theta\in\mathds{R}}\Re\bra{\psi_0}e^{i\theta}\ket{\psi_1}) \\
	&= \arccos|\braket{\psi_0}{\psi_1}| \\
	&= \arccos\sqrt{\tr(\rho_0\rho_1)}.
\end{split}
\end{equation}
This is the Fubini-Study distance \eqref{FSdistance}.

\begin{rmk}\label{Stiefels}
The Hopf bundle can be generalized to a bundle over $G(n,\HH)$ for a general $n$. The generalization is called the Stiefel bundle of $n$-frames. Edelman et al.\ \cite{EdArSm1998} used Stiefel bundles to derive the formula \eqref{diastnceGII} for the Grassmann distance. 
\end{rmk}

\subsection{Preparation for the Uhlmann bundle}
Before we define the Uhlmann bundle we need to introduce some more notation. Let $n$ be a positive integer not greater than the dimension of $\HH$. Fix an $n$-dimensional subspace $\HH^n$ of $\HH$, and let $B(n,\HH)$ be the space of linear operators from $\HH^n$ to $\HH$ equipped with the Hilbert-Schmidt Hermitian product, $\braket{B_0}{B_1}=\tr(B_0^\dagger B_1)$.\footnote{No result will depend on the choice of subspace $\HH^n$.} The space $B(n,\HH)$ is a parallelizable manifold whose tangent spaces can be canonically identified with $B(n,\HH)$ regarded as a real vector space. We equip $B(n,\HH)$ with the translation-invariant metric $\gHS$ that agrees with the real part of the Hilbert-Schmidt product at the origin,
\begin{equation}
	\gHS(\dot B_a,\dot B_b)
	= \frac{1}{2}\tr(\dot B_a^\dagger\dot B_b+\dot B_b^\dagger\dot B_a).
\end{equation}

The Uhlmann bundle is a principal fiber bundle over the space of density operators on $\HH$ of rank $n$. We denote this space by $S(n,\HH)$. The space $S(n,\HH)$ is a non-compact manifold if $n>1$, and equals $P(\HH)$ if $n=1$. The total space of the Uhlmann bundle is $W(n,\HH)$, consisting of all operators $W$ in $B(n,\HH)$ such that $W^\dagger W$ has rank $n$ and unit trace. The symmetry group is the Lie group $\U(n)$ of unitary operators on $\HH^n$. The symmetry group acts on $W(n,\HH)$ from the right by operator composition. The Lie algebra of $\U(n)$ is the Lie algebra $\uu(n)$ of skew-Hermitian operators on $\HH^n$.

\subsection{The Uhlmann bundle}\label{Uhlmannbundle}
The Uhlmann bundle is the $\U(n)$-principal bundle $\wp_n$ from $W(n,\HH)$ onto $S(n,\HH)$ defined as $\wp_n(W)=WW^\dagger$.  The Uhlmann bundle is a generalization of the Hopf bundle, which we recover for $n=1$. Following Uhlmann we call the elements of $W(n,\HH)$ amplitudes for the density operators in $S(n,\HH)$.

The vertical bundle is the bundle of kernels of the differential of $\wp_n$. That a tangent vector $\dot W$ at $W$ is vertical is equivalent to $\dot W W^\dagger+ W \dot W^\dagger=0$. We equip $W(n,\HH)$ with the restriction $\gW$ of the metric $\gHS$, and we define the horizontal bundle as the orthogonal complement of the vertical bundle with respect to $\gW$. The vertical space at $W$ consists of all operators of the form $WX$ where $X$ belongs to $\uu(n)$, and the Uhlmann connection form is the $\uu(n)$-valued $1$-form $\AU$ defined by the assumption that for every tangent vector $\dot W$ at $W$, the vector $W\!\AU(\dot W)$ is the vertical projection of $\dot W$. H{\"u}bner \cite{Hu1993} derived an explicit formula for Uhlmann's connection form. We will not need this formula here.

By definition, a tangent vector $\dot W$ at $W$ is horizontal if $\gW(\dot W,WX)=0$ for every skew-Hermitian operator $X$ on $\HH^n$. This condition is equivalent to 
\begin{equation}\label{horizcond}
	\dot W^\dagger W=W^\dagger \dot W.
\end{equation}
Since the symmetry group acts by isometries on $W(n,\HH)$, we can project $\gW$ to a metric $\gB$ on $S(n,\HH)$. The metric $\gB$ is defined as follows. Suppose that $\dot\rho_a$ and $\dot\rho_b$ are tangent vectors at $\rho$. Pick any amplitude $W$ of $\rho$ and let $\dot W_a$ and $\dot W_b$ be the horizontal lifts of $\dot\rho_a$ and $\dot\rho_b$ at $W$. Then $\gB(\dot\rho_a,\dot\rho_b)=\gW(\dot W_a,\dot W_b)$. The metric $\gB$ is known as the Bures metric \cite{Bu1969}. See also \cite{Jo1994,CrUh2009}. The Bures metric is proportional to the quantum Fisher information; see Remark \ref{Frowis} below. The geodesic distance function associated with the Bures metric is called the Bures angle.\footnote{The Bures angle should not be confused with the Bures distance which is the geodesic distance associated with $\gB$'s extension to the space of positive operators on $\HH$.}

\subsection{Uhlmann's energy dispersion estimate}\label{Uhlmanns estimate}
Consider a curve of density operators $\rho_t$ in $S(n,\HH)$. Let $W_t$ be a lift of $\rho_t$, and let $X_t=\AU(\dot W_t)$. Then
\begin{equation}
	\|\dot\rho_t\|^2_{\gB} 
	= \|\dot W_t\|^2_{\gW} - \|W_tX_t\|^2_{\gW}.
\end{equation}
If $\rho_t$ is generated by $H_t$, then $W_t$ can be chosen such that $\dot W_t=-iH_tW_t$. The expectation value of $H_t$ at $\rho_t$ equals the magnitude of the orthogonal projection of $W_tX_t$ on the vertical unit field $-iW_t$:
\begin{equation}\label{niklas}
\begin{split}
    \tr(H_t\rho_t)
	&= \gW(-iW_t,-iH_tW_t) \\
	&= \gW(-iW_t,W_tX_t).
\end{split}
\end{equation}
Write $W_tY_t$ for the orthogonal projection of $W_tX_t$ on the hyperplane perpendicular to $-iW_t$ in the vertical space at $W_t$. By \eqref{niklas}, the magnitude of $W_tY_t$ squared is
\begin{equation}
	\|W_tY_t\|^2_{\gW}
	= \|W_tX_t\|^2_{\gW} - \tr(H_t\rho_t)^2.
\end{equation}
From this observation follows that
\begin{equation}\label{just}
\begin{split}
	\|\dot\rho_t\|^2_{\gB}
	&= \|\dot W_t\|^2_{\gW} - \|W_tX_t\|^2_{\gW} \\
	&= \tr(H_t^2\rho_t) - \|W_tX_t\|^2_{\gW} \\
	&= \Delta^2(H_t,\rho_t) - \|W_tY_t\|^2_{\gW}.
\end{split}
\end{equation}
The size of $W_tY_t$ thus determines how much the speed of $\rho_t$ deviates from the energy uncertainty. For $n=1$, the vertical bundle is $1$-dimensional and, hence, $W_tY_t=0$. This observation explains equation \eqref{FSvariance} geometrically. At the same time, \eqref{just} indicates that equation \eqref{FSvariance} seldom holds for the Bures metric if the state is mixed. We will return to this issue in Section \ref{Utight}.

According to equation \eqref{just}, the square of the Bures speed is less than or equal to the energy variance. Consequently, $\Delta t\Delta E$ is greater than or equal to the Bures angle between $\rho_0$ and $\rho_1$. The inequality 
\begin{equation}\label{Uhqsl}
	\Delta t \Delta E \geq \distB(\rho_0,\rho_1)
\end{equation}
is the Uhlmann energy dispersion estimate \cite{Uh1992b}. Unlike Uhlmann's, the derivation above does not rely on the assumption that the initial state is faithful. See also \cite{JoKo2010,DeLu2013} for alternative derivations of the estimate \eqref{Uhqsl}.

\subsection{The Bures angle}
The amplitude space $W(n,\HH)$ is an open and dense subset of the unit sphere in $B(n,\HH)$. Hence, geodesics in $W(n,\HH)$ are great spherical arcs. Bures geodesics lift to horizontal great arcs, and a shortest geodesic between $\rho_0$ and $\rho_1$ lift to a shortest great arc connecting the fibers over $\rho_0$ and $\rho_1$. The spherical distance between any two amplitudes $W_0$ and $W_1$ in the fibers over $\rho_0$ and $\rho_1$ is $ \arccos \Re\tr(W_0^\dagger W_1)$. Consequently,
\begin{equation}\label{call}
\begin{split}
	\distB(\rho_0,\rho_1)
	&= \min_{U\in\U(n)}\arccos\Re\tr(W_0^\dagger W_1 U) \\
	&= \arccos\Big(\max_{U\in\U(n)}\Re\tr(W_0^\dagger W_1 U)\Big) \\
	&= \arccos\tr\big|W_0^\dagger W_1\big|.
\end{split}
\end{equation}
This calculation generalizes that in equation \eqref{calc}. The last identity is well known \cite{NiCh2010}, but for convenience we have included a proof in Appendix \ref{Uhlmanns_theorem}. There we also show that $\tr|W_0^\dagger W_1| = \tr|\sqrt{\rho_0}\sqrt{\rho_1}\,|$. The quantity $\tr|\sqrt{\rho_0}\sqrt{\rho_1}\,|$ is the square root of the fidelity between $\rho_0$ and $\rho_1$; see \cite{Uh1976}. 

We can rewrite Uhlmann's dispersion estimate as 
\begin{equation}\label{Buresavstand}
	\Delta t\geq \frac{ \arccos\tr|\sqrt{\rho_0}\sqrt{\rho_1}\,|}{\Delta E}.
\end{equation}
The expression on the right is the Uhlmann QSL, which we denote by $\tauU$. The Uhlmann QSL is one of the most widely used extensions of the Mandelstam-Tamm QSL. In the next section we investigate its tightness. 

\begin{ex}\label{ex:GrassmanvsUhlmann}
Assume, as in Example \ref{ex:non-degenerate-comm}, that $\rho_0$ and $\rho_1$ are isospectral, commuting, and nondegenerate mixed states with spectral decompositions
\begin{align}
	\rho_0 &= \sum_{j=1}^N p_j\ketbra{u_j}{u_j},\\
	\rho_1 &= \sum_{j=1}^N p_{\sigma(j)}\ketbra{u_{j}}{u_{j}}.
\end{align} 
Then
\begin{align}
	\tauU&=\frac{1}{\Delta E} \arccos\bigg(\sum_{j=1}^N \sqrt{p_jp_{\sigma(j)}}\bigg),\label{mint}\\
	\tauG&=\frac{\pi}{2\Delta E}\sqrt{ \sum_{j\ne \sigma(j)} p_j }.\label{starts}
\end{align}
Using that $\arccos^2 x\leq \pi^2(1-x)/4$ for $0\leq x\leq 1$, with strict inequality for $0 < x < 1$, we get that
\begin{equation}
\begin{split}
	\arccos^2\!\bigg(\sum_{j=1}^N \sqrt{p_jp_{\sigma(j)}}\bigg)
	&\leq \frac{\pi^2}{4}\bigg(1-\sum_{j=1}^N \sqrt{p_jp_{\sigma(j)}}\bigg) \\
	&= \frac{\pi^2}{4}\hspace{-3pt}\sum_{j\ne \sigma(j)}\hspace{-4pt}\big(p_j-\sqrt{p_jp_{\sigma(j)}}\big) \\
	&\leq \frac{\pi^2}{4}\hspace{-3pt}\sum_{j\ne \sigma(j)}\hspace{-4pt} p_j.
\end{split}
\end{equation}
Since the states are assumed to be mixed, that is, have a rank greater than two, and nondegenerate, the first or the last inequality is strict. Consequently, $\tauG>\tauU$.
\end{ex}

In the next section we will prove that $\tauU$ is never tight for systems in faithful states. As we saw in Example \ref{ex:two-cycles}, this is not the case for $\tauG$; there are faithful states that can be transformed one into the other in such a way that $\tauG$ equals the evolution time. 

\subsection{On the tightness of the Uhlmann QSL}\label{Utight}
Let $\rho_t$, $W_t$, and $X_t$ be as in Section \ref{Uhlmanns estimate}. The curve $W_t$ can be adjusted fiberwise to a horizontal lift of $\rho_t$:
\begin{equation}\label{denna}
	W_t^h=W_t\,\texp\Big(\!-\!\int_{t_0}^t\dt\, X_t \Big).
\end{equation}
The ``$\mathrm{T}$'' preceding the exponential is the positive time-ordering operator. That $W_t^h$ satisfies the horizontality condition \eqref{horizcond} is equivalent to 
\begin{equation}\label{hoho}
	 W_t^\dagger \dot W_t - \dot W_t^\dagger  W_t 
	 = \{X_t,W_t^\dagger W_t\},
\end{equation}
where $\{\cdot,\cdot\}$ is the anti-commutator. Moreover, the Bures speed of $\rho_t$ is $\|\dot\rho_t\|_{\gB} = \|\dot W_t^h\|_{\gW}$.

Equation \eqref{just} shows that a necessary condition for the estimate \eqref{Uhqsl} to be an identity is that $W_tX_t$ is proportional to $-iW_t$. That is, $X_t = -i\alpha_t\1$ for some real-valued $\alpha_t$. If such is the case, equation \eqref{hoho} reduces to 
\begin{equation}\label{abel}
	W_t^\dagger H_t W_t = \alpha_tW_t^\dagger W_t.
\end{equation}
Since $W_t$ defines a bijection between $\HH^n$ and the support of $\rho_t$, equation \eqref{abel} implies that if $\Pi_t$ is the projector onto the support of $\rho_t$, then 
\begin{equation}
	\Pi_t H_t \Pi_t = \alpha_t\Pi_t.
\end{equation}
This shows that for systems in a faithful state, in which case $\Pi_t$ is the identity operator, the Uhlmann QSL is not tight. Below we will derive a more general result covering systems in faithful states as a special case.

Next we consider systems for which the state's rank is at most half of the dimension of the Hilbert space. For such systems there are Hamiltonians generating Bures geodesics. To see this, let $\Pi_0$ be the orthogonal projection onto the support of $\rho_0$, and take any time-independent Hamiltonian $H$ satisfying the two conditions 
\begin{align}
	&\Pi_0 H \Pi_0 + (\1-\Pi_0) H (\1-\Pi_0) = 0, \label{i}\\
	&\Pi_0 H (\1-\Pi_0) H \Pi_0 = \beta\Pi_0, \label{ii}
\end{align}
where $\beta$ is some positive real number. Consider $\rho_t$ generated by $H$ and extending from $\rho_0$, and let $W_t$ be a  lift of $\rho_t$ such that $\dot W_t=-iHW_t$. Since the support of $W_0$ is the same as that of $\rho_0$, condition \eqref{i} ensures that $W_t$ is horizontal:
\begin{equation}
\begin{split}
	\dot W^\dagger_t W_t - W^\dagger_t \dot W_t
	&= 2i W^\dagger_t H W_t \\
	&= 2i W^\dagger_0 H W_0 \\
	&= 2i W^\dagger_0\Pi_0 H \Pi_0 W_0 \\
	&= -2i W^\dagger_0(\1-\Pi_0) H (\1-\Pi_0) W_0 \\
	&= 0.
\end{split}
\end{equation}
By \eqref{just}, the Bures speed of $\rho_t$ equals the energy uncertainty. Furthermore, conditions \eqref{i} and \eqref{ii} together ensure that $W_t$ is a great arc and, hence, that $\rho_t$ is a geodesic. To see this, first note that
\begin{equation}
\begin{split}
	H^2 
	&= \big(\Pi_0 H (\1-\Pi_0) + (\1-\Pi_0) H \Pi_0\big)^2 \\ 
	&= \beta \Pi_0 + (\1-\Pi_0) H \Pi_0 H (\1-\Pi_0).
\end{split}
\end{equation}
Then, if $U_t$ is the time-evolution operator of $H$, 
\begin{equation}
\begin{split}
	\ddot W_t 
	&= -H^2 W_t \\
	&= -U_t H^2 W_0 \\
	&= -U_t \big(\beta \Pi_0 + (\1-\Pi_0) H \Pi_0 H (\1-\Pi_0)\big) W_0 \\
	&= -U_t \beta \Pi_0 W_0 \\
	&= -\beta W_t.
\end{split}
\end{equation}
Since geodesics are local length minimizers, the Uhlmann dispersion estimate \eqref{Uhqsl} is saturated for small evolution time intervals. Note, however, that this does not mean that Uhlmann's QSL is tight if only the geodesic distance between initial and final state is small enough; we have only shown that $\tauU$ is tight for some nearby states.

If $\rho_0$'s rank is strictly greater than half of the dimension of $\HH$, condition \eqref{ii} cannot be fulfilled; there are no Hermitian operators and positive $\beta$ for which $\eqref{ii}$ holds. Next we show that if by assumption $H_t$ generates a locally shortest curve in $S(n,\HH)$ having a speed equal to the energy uncertainty, then $H_t$ has to satisfy the impossible condition in \eqref{ii}, possibly after some legitimate modification. This contradiction lets us conclude that the Uhlmann QSL is not tight even for short time intervals. 

Assume that $H_t$ generates a locally shortest curve $\rho_t$ having a Bures speed equal to the energy uncertainty. In Appendix \ref{horizontal_modification} we show how to modify $H_t$ so that $\rho_t$ becomes reparameterized to a geodesic and $W_t$ becomes reparameterized and fiberwise adjusted to a horizontal geodesic. Assuming this, the extrinsic acceleration $\ddot W_t$, that is, $W_t$'s covariant derivative as a curve in $B(n,\HH)$, is parallel to $W_t$. This is equivalent to 
\begin{equation}\label{ops}
	i\dot H_tW_t+H_t^2W_t=\beta_t W_t
\end{equation} 
for some real-valued $\beta_t$. Let $\Pi_t$ be the projection onto the support of $\rho_t$. According to the second paragraph in the current section, $\Pi_tH_t\Pi_t=0$. But then also $\Pi_t\dot H_t\Pi_t=0$. The amplitude $W_t$ has the same support as $\rho_t$. Hence $W_t=\Pi_tW_t$. From \eqref{ops} follows that 
\begin{equation}\label{oops}
\begin{split}
	\beta_t \Pi_t W_t
	&= i\Pi_t\dot H_t\Pi_t W_t+\Pi_t H_t^2\Pi_t W_t \\
	&=\Pi_t H_t^2\Pi_t W_t,    
\end{split}
\end{equation} 
which implies that
\begin{equation}\label{ooops}
\begin{split}
	\beta_t \Pi_t 
	&= \Pi_t H_t^2\Pi_t \\
	&= \Pi_t H_t(\1-\Pi_t)H_t\Pi_t.
\end{split}
\end{equation} 
This equation can be satisfied only if $\beta_t=0$. But then $H_t$ commutes with $\rho_t$, implying that $\rho_t$ is stationary.

The discussion above raises the question of whether there is a metric such that the speed of a curve generated by a Hamiltonian is equal to the energy uncertainty, regardless of the state's rank. We will prove in Section \ref{Unitorbits} that the answer is ``No.'' However, we will also construct a metric that provides the tightest general extension of the Mandelstam-Tamm QSL. 

\begin{rmk}\label{Frowis}
Fr{\"o}wis \cite{Fr2012} suggested replacing the energy uncertainty with the quantum Fisher information 
\begin{equation}
	\F(H,\rho)=2\sum_{j,k=1}^m \frac{(p_j-p_k)^2}{p_j+p_k}\tr(HP_j HP_k)
\end{equation}
in the Uhlmann QSL when studying the effect of entanglement on the evolution time. Quantum Fisher information is proportional to the Bures' speed squared,
\begin{equation}
	\F(H,\rho)=4\|-i[H,\rho]\|^2_{\gB}.
\end{equation}
Hence, by \eqref{just},
\begin{equation}
	\Delta t
	\geq \frac{2\arccos\tr |\sqrt{\rho_0}\sqrt{\rho_1}\,|}{\big\langle\! \sqrt{\F(H_t,\rho_t)}\,\big\rangle}.
\end{equation}
The bound on the right is the Fr{\"o}wis QSL. 

The Fr{\"o}wis QSL is always greater than the Uhlmann QSL. Whether the Fr{\"o}wis QSL is tight or not for systems in mixed states was formulated as an open question in the review article \cite{Fr2016}. A careful examination of the arguments underlying the discussion in Section \ref{Utight} shows that if the rank of the state exceeds half of the Hilbert space dimension, the Fr{\"o}wis QSL is not tight; if the rank is less than or equal to half of the Hilbert space dimension, there are states for which the Fr{\"o}wis QSL is tight. However, the authors believe that the latter is a non-generic situation if the rank is greater than one.
\end{rmk}

\begin{rmk}\label{GrassmannvsUhlmann}
In Section \ref{Utight} we showed that if the rank $n$ of the state is high enough, the Uhlmann QSL is not tight. This is because geodesics of the Bures metric on the manifold of states having rank $n$ are not generated by Hamiltonians. One way to improve Uhlmanns QSL is to restrict the Bures metric to the manifold of states having the same spectrum as the initial state, and then replace the Bures angle with the distance function of the restricted metric. This would lead to a tighter extension of the Mandelstam-Tamm QSL than Uhlmann's QSL. 

In Section \ref{ReltoG} we will show how to pull back the product metric in \eqref{Grassmannprodmetric} to the manifold of states isospectral to the initial state. That this will lead to an extension of the Mandelstam-Tamm QSL that is even stronger than the one obtained from the restriction of the Bures metric follows from the quantity $J(H,\rho)$, defined in \eqref{jay}, being greater than a quarter of the Fisher information: 
\begin{equation}\label{jayexceedsF}
	J(H,\rho) 
	\geq \frac{1}{4}\F(H,\rho).
\end{equation}
Notice that this inequality does not follow immediately from \eqref{duo} since a quarter of the Fisher information is greater than the Wigner-Yanase skew information; see \cite{Lu2004}. To prove the inequality \eqref{jayexceedsF}, we first observe that for any pair of eigenvalues,
\begin{equation}
	(p_j+p_k)(1-\delta_{jk})\geq \frac{(p_j-p_k)^2}{p_j+p_k},
\end{equation}
where $\delta_{jk}$ is the Kronecker delta. A simple rewriting of $J(H,\rho)$ then yields the inequality \eqref{jayexceedsF}:
\begin{equation}
\begin{split} 
	J(H&,\rho) 
	= \sum_{j=1}^m p_j \big (\tr(H^2P_j)-\tr(HP_jHP_j)\big) \\
	&= \frac{1}{2}\sum_{j=1}^m \sum_{k=1}^m(p_j+p_k)(1-\delta_{jk})\tr(HP_jHP_k)\\
	&\geq \frac{1}{2}\sum_{j=1}^m\sum_{k=1}^m \frac{(p_j-p_k)^2}{p_j+p_k}\tr(HP_jHP_k) \\
	&= \frac{1}{4}\F(H,\rho).
\end{split}
\end{equation}
\end{rmk}

\section{A Wigner-Yanase QSL}\label{Wigge}
We saw in the previous section that Uhlmann's energy dispersion estimate \eqref{Uhqsl} is never saturated if the state is faithful. In this section we show that if the eigenvalue spectrum of the faithful state has a width not exceeding three times the lowest eigenvalue, then two times the skew information is upper bounded by the energy variance. Two times the skew information is also the square of the evolving state's speed with respect to a metric whose associated geodesic distance always is greater than the Bures distance. This observation results in a tighter and easier to compute QSL than that of Uhlmann.

\subsection{The square root section}
Consider the Uhlmann bundle for faithful states. Let $\zeta$ be the section assigning the square root to each state, $\zeta(\rho)=\sqrt{\rho}$, and let $\gWY$ be the pull-back of $\gW$ by $\zeta$,
\begin{equation}
	\gWY(\dot\rho_a,\dot\rho_b)= \gW(d\zeta(\dot\rho_a),d\zeta(\dot\rho_b)).
\end{equation}
The metric $\gWY$ dominates the Bures metric since $\gWY$ assigns the size of $d\zeta(\dot\rho)$ to the size of $\dot\rho$ while $\gB$ assigns the size of the horizontal projection of $d\zeta(\dot\rho)$ to the size of $\dot\rho$. The geodesic distance associated with $\gWY$ is therefore always greater than the Bures angle, 
\begin{equation}\label{WYbiggerthanB}
	\distWY(\rho_0,\rho_1) \geq \distB(\rho_0,\rho_1).
\end{equation}

Consider a $\rho_t$ generated by a Hamiltonian $H_t$. Write $\rho_0$ and $\rho_1$ for the initial and the final state. The push-forward of the velocity field of $\rho_t$ by $\zeta$ is $d\zeta(\dot\rho_t)=-i[H_t,\sqrt{\rho_t}\,]$ and, consequently, the speed of $\rho_t$ is the square root of twice the Wigner-Yanase skew information:\footnote{This is the reason why the metric has the subscript ``WY''.}
\begin{equation}\label{yanne}
	\|\dot\rho_t\|^2_{\gWY}
	= \tr\big((-i[H_t,\sqrt{\rho_t}\,])^2\big)
	= 2I(H_t,\rho_t).
\end{equation}
It follows that 
\begin{equation}\label{tjabor}
	\Delta t \big\langle\! \sqrt{2I(H_t,\rho_t)}\,\big\rangle 
	\geq \distWY(\rho_0,\rho_1).
\end{equation}
Below we show that if the width of the spectrum of $\rho_0$, that is, the difference between the greatest and the smallest eigenvalue of $\rho_0$, is smaller than three times the lowest eigenvalue, then $2I(H_t,\rho_t)\leq \Delta^2(H_t,\rho_t)$. Hence, for such systems,
\begin{equation}\label{theWYqsl}
	\Delta t \geq \frac{\distWY(\rho_0,\rho_1)}{\Delta E}.
\end{equation}

Let $P_{j;t}$ be the projector onto the $j$th eigenspace of $\rho_t$ and define the horizontal part of $H_t$ as\footnote{The terminology will be explained in Section \ref{isospektral}.}
\begin{equation}
	H^h_{t}=\sum_{j\ne k} P_{j;t} H_t P_{k;t}.
\end{equation}
The horizontal part also generates $\rho_t$. Furthermore,
\begin{align}
	&I(H^h_{t},\rho_t) = I(H_t,\rho_t), \\
	&\Delta^2(H^h_{t},\rho_t) \leq \Delta^2(H_t,\rho_t).
\end{align}
It is thus sufficient to show that two times the Wigner-Yanase skew information of $\rho_t$ relative $H^h_{t}$ is less than the variance of $H^h_{t}$ at $\rho_t$. The expectation value of $H^h_{t}$ at $\rho_t$ vanishes and, hence,
\begin{equation}\label{tufft}
\begin{split} 
	\Delta^2(H^h_{t}&,\rho_t) - 2I(H^h_{t},\rho_t) \\
	&= 2\tr (H^h_{t}\sqrt{\rho_t}H^h_{t}\sqrt{\rho_t}) - \tr((H^h_{t})^2\rho_t) \\
	&= \sum_{j,k=1}^m (2\sqrt{p_jp_k} - p_j)\tr(H^h_tP_{j;t}H^h_tP_{k;t}). 
\end{split}
\end{equation}
Consequently, $2I(H_t,\rho_t)\leq \Delta^2(H_t,\rho_t)$ if $p_j \leq 2\sqrt{p_jp_k}$ for all $j$ and $k$. This is the case if and only if the width of the spectrum of $\rho_0$ is smaller or equal to three times its smallest eigenvalue.

\begin{rmk}
From equations \eqref{duo} and \eqref{tjabor} it follows that
\begin{equation}
	\Delta t\geq \frac{\distWY(\rho_0,\rho_1)}{\big\langle\! \sqrt{2I(H_t,\rho_t)}\,\big\rangle }\geq \frac{\distWY(\rho_0,\rho_1)}{\sqrt{2}\Delta E}.
\end{equation}
See also \cite{PiCiCeAdSo-Pi2016}. The calculation in \eqref{tufft} sharpens the second estimate under the given spectral condition.
\end{rmk}

\subsection{The Wigner-Yanase distance}
By construction, the square root section $\zeta$ is an isometric embedding of $S(n,\HH)$ with metric $\gWY$ in $W(n,\HH)$. Furthermore, as was first observed by Gibilisco and Isola \cite{GiIs2003}, every great arc in $W(n,\HH)$ which connects two elements in the image of $\zeta$ is completely contained in the image of $\zeta$. To see this, take any two faithful density operators $\rho_0$ and $\rho_1$. The great arc that connects $\sqrt{\rho_0}$ and $\sqrt{\rho_1}$ is, up to reparameterization, 
\begin{equation}
	W_t=
	\frac{ (1-t)\sqrt{\rho_0} + t\sqrt{\rho_1} }{ \|(1-t)\sqrt{\rho_0} + t\sqrt{\rho_1}\|_{\textsc{hs}} },
\end{equation}
where $0\leq t\leq 1$ and $\|\cdot\|_{\textsc{hs}}$ is the norm associated with the Hilbert-Schmidt product. Every $W_t$ is a positive operator, and $\rho_t=W_tW_t^\dagger$ is faithful. Hence, $W_t=\sqrt{\rho_t}$.

From the observation by Gibilisco and Isola follows that the Wigner-Yanase distance between $\rho_0$ and $\rho_1$ equals the spherical distance between $\sqrt{\rho_0}$ and $\sqrt{\rho_1}$,
\begin{equation}
	\distWY(\rho_0,\rho_1) 
	= \arccos\tr\big(\sqrt{\rho_0}\sqrt{\rho_1}\,\big).
\end{equation}
The quantity $\tr(\sqrt{\rho_0}\sqrt{\rho_1}\,)$ is called the affinity of $\rho_0$ and $\rho_1$; see \cite{LuZh2004}. The affinity is always less than the square root of the fidelity,
\begin{equation}
	\tr(\sqrt{\rho_0}\sqrt{\rho_1}\,)
	\leq \tr |\sqrt{\rho_0}\sqrt{\rho_1}\,|.
\end{equation}
Since $\arccos$ is decreasing, this observation corroborates that the Wigner-Yanase distance is greater than the Bures angle, as does the main result in \cite{Lu2004}. We write $\tauWY$ for the lower bound in \eqref{theWYqsl}, 
\begin{equation}
	\tauWY =\frac{\arccos\tr\big(\sqrt{\rho_0}\sqrt{\rho_1}\,\big)}{\Delta E}.
\end{equation}

Although $\tauWY$ is well defined for states of all ranks, $\tauWY$ need not be a QSL. It certainly is not for pure states unless the initial and final states are perpendicular, and need not be so for faithful states either as Example \ref{motes} shows. As was proven above, the condition
\begin{equation}\label{cravat}
	\width(\spec\rho_0) \leq 3\min(\spec\rho_0)
\end{equation}
guarantees that $\tauWY$ is a QSL.

\begin{ex}\label{WYbetterthanU}
Consider the isospectral qubit states
\begin{align}
	\rho_0 &= p\ketbra{0}{0}+(1-p)\ketbra{1}{1},\\
	\rho_1 &= p\ketbra{+}{+}+(1-p)\ketbra{-}{-},
\end{align}
where $\frac{1}{2} < p\leq \frac{4}{5}$ and $\ket{\pm}=\frac{1}{\sqrt{2}}(\ket{0} \pm \ket{1})$. We have that 
\begin{align}
	&\width(\spec\rho_0) = 2p-1, \\
	&\min(\spec\rho_0) = 1-p.
\end{align}
Since $2p-1\leq 3(1-p)$, the condition in \eqref{cravat} is fulfilled. The Wigner-Yanase and Bures geodesic distances are
\begin{align}
	&\distWY(\rho_0,\rho_1)=\arccos\Big(\frac{1}{2}+\sqrt{p(1-p)}\Big),\\
	&\distB(\rho_0,\rho_1)=\arccos\sqrt{\frac{1}{2}+2p(1-p)}.
\end{align}
Since $\arccos$ is strictly decreasing and
\begin{equation}
    \frac{1}{2}+\sqrt{p(1-p)} 
    < \sqrt{\frac{1}{2}+2p(1-p)},
\end{equation}
$\tauWY$ is strictly greater than $\tauU$.
\end{ex}

The following example shows that if the condition in \eqref{cravat} is violated, then $\tauWY$ need not be a QSL.

\begin{ex}\label{motes}
Suppose that $p$ in the previous example exceeds $\frac{1}{2}\big(1+(2\sqrt{2}-2)^{1/2}\,\big)\approx 0.96$. Then the condition in \eqref{cravat} is \emph{not} satisfied, and we have that
\begin{equation}
     \tauWY 
     = \frac{1}{\Delta E}\arccos\Big(\frac{1}{2}+\sqrt{p(1-p)} \Big)
     > \frac{\pi}{4\Delta E}.
\end{equation}
The Hamiltonian $H=i(\ketbra{1}{0}-\ketbra{0}{1})$ transforms $\rho_0$ into $\rho_1$ in the time $\Delta t=\pi/4$, and $\Delta E=1$. Hence, in this case $\tauWY$ is not a QSL.
\end{ex}

\begin{rmk}
According to Example \ref{two}, the Grassmann QSL is tight for any pair of qubits. For the qubits in Examples \ref{WYbetterthanU} and \ref{motes}, equation \eqref{tusk} yields $\tauG=\pi/4\Delta E$.
\end{rmk}

\section{Extensions of the Mandelstam-Tamm QSL from geometries on spaces of isospectral states}\label{Unitorbits}
In this section we address the question of whether there is a `best' metric producing the tightest extension of the Mandelstam-Tamm QSL. Since the von Neumann equation is eigenvalue spectrum preserving, such a metric only needs to be defined on the space of states isospectral to the initial state, that is, on the unitary orbit of the initial state. The Bures and the Wigner-Yanase metric considered in Sections \ref{Ulam} and \ref{Wigge} are defined on the much larger manifold consisting of all the states having the same rank as the initial state. The spectrum of the initial state is typically not preserved along the geodesics of these metrics. Hence the non-tightness of the Uhlmann and Wigner-Yanase QSLs.

Since the geodesic distance associated with a metric depends not only on the metric but also on the manifold,\footnote{Geodesics of a metric restricted to a submanifold need not be geodesics in the ambient manifold.} one can derive stronger QSLs from the restrictions of the Bures and Wigner-Yanase metrics to the unitary orbit of the initial state; cf.\ the discussion in Remark \ref{GrassmannvsUhlmann}. However, as we will see, the tightest extension of the Mandelstam-Tamm QSL does not result from either of these restrictions. 

\subsection{Horizontal and vertical Hermitian operators}\label{isospektral}
Each solution to a von Neumann equation remains in the space of states isospectral to the initial state. Conversely, any two isospectral states can be connected by a solution to a von Neumann equation. Let $\p$ be an eigenvalue spectrum, and let $S(\p,\HH)$ be the space of density operators on $\HH$ with spectrum $\p$. The group of unitary operators on $\HH$ acts transitively on $S(\p,\HH)$ by left conjugation, $U\cdot\rho=U\rho \,U^\dagger$. Thus, $S(\p,\HH)$ is the unitary orbit of each of its members.\footnote{In particular, $S(\p,\HH)$ is a compact manifold.} Since the unitary group acts transitively on $S(\p,\HH)$, each tangent vector of $S(\p,\HH)$ at a $\rho$ can be written as $-i[A,\rho]$ for some Hermitian operator $A$. Notice that two Hermitian operators can represent the same tangent vector: $[A_1,\rho]=[A_2,\rho]$ if and only if $P_j A_1 P_k=P_j A_2 P_k$ for every pair of different eigenspace projectors $P_j$ and $P_k$ of $\rho$. This observation proposes a division of $A$ into a vertical and a horizontal component (the terminology is explained in Section \ref{bundleapproach}):
\begin{align}
	A^v &= \sum_{j=1}^{m} P_j A P_j,\\
	A^h &= A - A^v.
\end{align}
Two Hermitian operators then represent the same tangent vector at $\rho$ if and only if they have the same horizontal component at $\rho$. Note that the division depends on the state and that the divisions of a Hermitian operator at two different states do not have to be the same. We say that $A$ is horizontal at $\rho$ if $A^v = 0$, and we say that $A$ is almost horizontal at $\rho$ if $A^v$ acts as a multiple of the identity on the support of $\rho$. 

The variance of $A$ satisfies the Pythagorean identity 
\begin{equation}
	\Delta^2(A,\rho)
	= \Delta^2(A^h,\rho) + \Delta^2(A^v,\rho).
\end{equation}
Furthermore, $\Delta(A,\rho) = \Delta(A^h,\rho)$ if and only if $A$ is almost horizontal at $\rho$. To show this we use that $A^v$ and $\rho$ are simultaneously diagonalizable. Let $n_j$ be the multiplicity of the $j$th nonzero eigenvalue $p_j$ of $\rho$ and choose a spanning set of orthonormal vectors $\ket{j;1}, \ket{j;2},\dots,\ket{j;n_j}$ in the eigenspace corresponding to $p_j$ that are also eigenvectors of $A^v$, say $A^v\ket{j;k}=a_{jk} \ket{j;k}$. Now $\Delta(A,\rho) = \Delta(A^h,\rho)$ if and only if the variance of $A^v$ vanishes at $\rho$, or equivalently if
\begin{equation}
	\sum_{j=1}^m\sum_{k=1}^{n_j} p_j a_{jk}^2
	= \Big(\sum_{j=1}^m\sum_{k=1}^{n_j} p_j a_{jk}\Big)^2.
\end{equation}
Since the squaring function is strictly convex, this identity is satisfied if and only if all the $a_{jk}$s are identical. 

\subsection{The metric}
Is there a metric on $S(\p,\HH)$ such that the speed always equals the energy uncertainty? By ``always'' we mean that for every Hamiltonian $H_t$, if $H_t$ generates $\rho_t$, then the speed at any time $t$ is $\Delta(H_t,\rho_t)$. Such a metric would give rise to the tightest possible quantum speed limit of the form in \eqref{genqsl}, with the numerator being the geodesic distance. We have seen that such a metric exists if $\p$ describes the spectrum of a pure state. In that case, $S(\p,\HH)$ is the projective Hilbert space, and the Fubini-Study metric is such that the speed is always equal to the energy uncertainty. However, if $\p$ describes the spectrum of a mixed state, there is no such metric. The reason is that two Hermitian operators with the same horizontal component at a $\rho$ do not have to have the same uncertainty at $\rho$. Since a Hermitian operator and its horizontal component represent the same tangent vector, we have that if a metric $g$ is such that $\|-i[A,\rho]\|_{g} \leq \Delta(A,\rho)$ for every $\rho$ and every $A$, the metric must also meet the stricter condition $\|-i[A,\rho]\|_{g} \leq \Delta(A^h,\rho)$. Interestingly, there is a metric that does and which saturates this latter condition, namely 
\begin{equation}\label{metrik}
	\gP\big(-i[A_1,\rho],-i[A_2,\rho]\big)
	= \frac{1}{2}\tr\big(\{A^h_1,A^h_2\}\rho\big).
\end{equation}
In Section \ref{bundleapproach} we relate $\gP$ to the Bures metric.

\subsection{The tightest general extension of the Mandelstam-Tamm QSL}\label{tightest} 
Consider a Hamiltonian $H_t$ that generates a curve $\rho_t$ in $S(\p,\HH)$ with spectral decomposition $\sum_{j=1}^m p_jP_{j;t}$. The vertical and horizontal components of $H_t$ at $\rho_t$ are 
\begin{align}
	H^v_t &= \sum_{j=1}^m P_{j;t} H_t P_{j;t}, \\
	H^h_t &= H_t - H^v_t.
\end{align}
The horizontal component is a Hamiltonian that also generates $\rho_t$, that is $\dot\rho_t=-i[H^h_t,\rho_t]$, and
\begin{equation}\label{not}
	\|\dot\rho_t\|_{\gP}
	=\Delta(H^h_t,\rho_t)\leq \Delta(H_t,\rho_t).
\end{equation}
We call $H_t$ parallel transporting if $H_t=H_t^h$ for every $t$, and we call $H_t$ almost parallel transporting if $H_t$ is almost horizontal at $\rho_t$ for every $t$, that is, if $H_t^v$ acts as a scalar on the support of $\rho_t$. By equation \eqref{not}, $\|\dot\rho_t\|_{\gP}=\Delta(H_t,\rho_t)$ if and only if $H_t$ is almost parallel transporting. If $\p$ is the spectrum of pure states, all Hamiltonians are almost parallel transporting, and $\gP$ is the Fubini-Study metric.

Let $\rho_0$ and $\rho_1$ be the initial and the final state of $\rho_t$. The estimate $\Delta t \Delta E \geq \dist_{\gP}(\rho_0,\rho_1)$ follows immediately from \eqref{not}, with the right-hand side being the geodesic distance between $\rho_0$ and $\rho_1$. We define the QSL $\tauP$ as
\begin{equation}\label{theqsl}
	\tauP = \frac{\dist_{\gP}(\rho_0,\rho_1)}{\Delta E}.
\end{equation}
For pure states, $\tauP=\tauMT$. Thus, $\tauP$ is an extension of the Mandelstam-Tamm QSL.

Assume that $d$ is a real-valued function that takes pairs of states, say $\rho_0$ and $\rho_1$, from $S(\p,\HH)$ as an argument and which is such that $\Delta t \Delta E\geq d(\rho_0,\rho_1)$ for every Hamiltonian that connects $\rho_0$ and $\rho_1$. Then $\Delta t \langle\Delta(H_t,\rho_t)\rangle \geq d(\rho_0,\rho_1)$ holds for every parallel transporting Hamiltonian $H_t$ connecting $\rho_0$ and $\rho_1$. Now, if $H_t$ is chosen such that $\rho_t$ is a shortest curve,\footnote{Such a Hamiltonian exists because any two states in $S(\p,\HH)$ can be connected by a shortest geodesic, $S(\p,\HH)$ being compact, and any such geodesic is generated by a Hamiltonian as $S(\p,\HH)$ is the unitary orbit of each of its states. The Hamiltonian can be made parallel transporting by removing its vertical component.} $\Delta t \langle\Delta(H_t,\rho_t)\rangle$ is the geodesic distance from $\rho_0$ to $\rho_1$. From this follows that $d(\rho_0,\rho_1)\leq \dist_{\gP}(\rho_0,\rho_1)$. Thus, in a sense, $\tauP$ is the tightest general extension of the Mandelstam-Tamm QSL. This QSL appeared for the first time in \cite{AnHe2014}.

\begin{rmk}
If the initial and the final state, as well as the evolutionary curves between them, are required to lie in a given submanifold of $S(\p,\HH)$, the distance function in $\tauP$ should be replaced by the geodesic distance associated with the restriction of $\gP$ to the submanifold. Then $\tauP$ will remain the tightest `general' extension of the Mandelstam-Tamm QSL.
\end{rmk}

\begin{rmk}
If the permitted Hamiltonians must meet certain conditions that exclude their horizontal components, then $\tauP$ no longer needs to be the tightest extension of the the Mandelstam-Tamm QSL, cf.\ \cite{BuSePo2019}.
\end{rmk}
	
\subsection{Relation to the Grassmann QSL}\label{ReltoG}
The quantity $J(A,\rho)$ defined in \eqref{jay} equals the variance of the horizontal component of $A$ at $\rho$. To see this, first note that $I(A,P_j)=I(A^h,P_j)$, which is a consequence of $[A^v,P_j]=0$ and $P_jA^hP_j=0$. Then, since the expectation value of $A^h$ at $\rho$ is zero,
\begin{equation}\label{sleepy}
	J(A,\rho)= J(A^h,\rho)=\tr((A^h)^2\rho)=\Delta^2(A^h,\rho).
\end{equation}

Consider the embedding $\iota$ of $S(\p,\HH)$ in $\prod_{j=1}^{\bar{m}}G(n_j,\HH)$ which sends a state to its $\bar{m}$-tuple of eigenspace projectors corresponding to the nonzero eigenvalues,
\begin{equation}
	\iota(\rho) = (P_1,P_2,\dots,P_{\bar{m}}).
\end{equation}
This embedding is isometric; if $A$ is horizontal at $\rho$, then
\begin{equation}
\begin{split}
	\|d\iota(-i[A,\rho])\|^2_{\gpG}
	&= J(A,\rho) \\
	&= \Delta^2(A,\rho) \\
	&= \|-i[A,\rho]\|^2_{\gP}.
\end{split}
\end{equation}
As a consequence, $\dist_{\gP}(\rho_0,\rho_1)\geq \distpG(\iota(\rho_0),\iota(\rho_1))$, which in turn implies that $\tauP\geq\tauG$. This observation corroborates that $\tauP$ is the tightest possible general extension of the Mandelstam-Tamm QSL. Example \ref{Grassmann_not_tight} shows that there are cases where $\tauP$ is strictly greater than $\tauG$.

\subsection{The bundle approach}\label{bundleapproach}
The metric $\gP$ is the projection of the restriction of $\gW$ defined in Section \ref{Uhlmannbundle} to the total space of a reduction of the Uhlmann bundle. We will here give a brief description of this fact. For a detailed account consult \cite{lic}.\footnote{Reference \cite{lic} also discusses the connection between the metric $\gP$ and the geometric phase introduced by Sj{\"o}qvist et al.\ \cite{SjPaEkAnErOiVe2000}.}

Let $n$ be the rank of the states with spectrum $\p$. Then $S(\p,\HH)$ is contained in $S(n,\HH)$. Let $\Lambda$ be a faithful density operator on $\HH^n$ having spectrum $\p$ with all the zero eigenvalues removed, if any, and let $W(\p,\HH)$ be the subspace of $W(n,\HH)$ consisting of all amplitudes $W$ such that $W^\dagger W=\Lambda$.\footnote{No result will depend on the choice of $\Lambda$; different choices give rise to isometric bundles \cite{lic}.} The restriction $\wpP$ of the Uhlmann projection $\wp_n$ to $W(\p,\HH)$ is a principal fiber bundle over $S(\p,\HH)$. The symmetry group of $\wpP$ is the subgroup of $\U(n)$ consisting of all the unitaries on $\HH^n$ that commutes with $\Lambda$. We equip $W(\p,\HH)$ with the restriction $\bgW$ of $\gW$. The vertical bundle of $\wpP$ is the kernel bundle of the differential of $\wpP$, and we define the horizontal bundle as the $\bgW$-orthogonal complement of the vertical bundle.

The group of unitary operators on $\HH$ acts from the left on $W(n,\HH)$ by operator composition, $U\cdot W = UW$. This action is invariant and transitive on $W(\p,\HH)$. Consequently, each tangent vector at a $W$ in $W(\p,\HH)$ has the form $-iAW$ for some Hermitian operator $A$. The differential of $\wpP$ sends $-iAW$ to $-i[A,\rho]$ where $\rho = WW^\dagger$. Let $A=A^h+A^v$ be the splitting of $A$ into a horizontal and a vertical component at $\rho$. Then $-iA^hW$ is the horizontal component and $-iA^vW$ is the vertical component of $-iAW$:
\begin{align}
	&d\wpP(-iA^vW)=-i[A^v,\rho]=0,\\
	&\bgW(-iA^hW,-iA^vW)=\frac{1}{2}\tr\big(\{A^h,A^v\}\rho\big)=0.
\end{align}
Since the symmetry group of $\wpP$ acts through isometries on $W(\p,\HH)$, we can project $\bgW$ to a metric $g$ on $S(\p,\HH)$; see \cite{Mo2006}. The projection is $\gP$:
\begin{equation}
\begin{split}
	g(-i[A_1,\rho],-i[A_2,\rho])
	&= \bgW(-iA^h_1W,-iA^h_2W) \\
	&= \frac{1}{2}\tr\big(\{A^h_1,A^h_2\}\rho\big) \\
	&= \gP(-i[A_1,\rho],-i[A_2,\rho]).
\end{split}
\end{equation}

\subsection{Hamiltonians generating geodesics}
The geodesic distance function associated with $\gP$ appears in the numerator of the QSL in \eqref{theqsl}, and, naturally, one would like to have an explicit formula for the geodesic distance function. Unfortunately, such a formula seems difficult to derive. One reason for this is that the geodesics of $\gP$ are generally generated by time-dependent Hamiltonians. This will be apparent from the geodesic equation that we derive in this section.

A related question is how to determine whether a given Hamiltonian generates a geodesic when it acts on a given state. To answer this question assume that $H_t$ generates a geodesic $\rho_t$ extending from the state $\rho_0$ with spectrum $\p$. According to a standard result from the theory of principal fiber bundles \cite{Mo2006}, $\rho_t$ is a geodesic if and only if all the horizontal lifts of $\rho_t$ are geodesics. As described in the third paragraph in Section \ref{bundleapproach}, any horizontal lift $W_t$ is a solution to the Schr{\"o}dinger equation $\dot W_t = -iH_t^hW_t$. Furthermore, $W_t$ is a geodesic if and only if its extrinsic acceleration $\ddot W_t$ is orthogonal to the tangent space of $W(\p,\HH)$ at $W_t$; see \cite{Sa1996}. Each vector in this tangent space can be written as $-iAW_t$ for a Hermitian operator $A$, and $\ddot W_t$ is orthogonal to $-iAW_t$ if 
\begin{equation}
\begin{split}
    \gHS(&\ddot W_t, -iAW_t)\\
    &= \frac{1}{2}\tr\bigg(\!\Big(\big(\dot H_t^h - i(H_t^h)^2\big)W_tW_t^\dagger \\
    &\hspace{59pt} + W_tW_t^\dagger\big(\dot H_t^h + i(H_t^h)^2\big)\Big)A\bigg) \\
	&=0.
\end{split}
\end{equation}
We conclude that $W_t$ is a geodesic if and only if
\begin{equation}
	\big(\dot H_t^h - i(H_t^h)^2\big)W_tW_t^\dagger + W_tW_t^\dagger\big(\dot H_t^h + i(H_t^h)^2\big)=0.
\end{equation}
Note that this equation is an equation for $\rho_t$. In summary, the Hamiltonian $H_t$ generates a geodesic $\rho_t$ if and only if the horizontal component $H^h_t$ of $H_t$ satisfies the equation
\begin{equation}\label{geodesiceq}
	\big(\dot H_t^h - i(H_t^h)^2\big)\rho_t + \rho_t\big(\dot H_t^h + i(H_t^h)^2\big)=0.
\end{equation}

At first glance it may seem that one needs to know the entire evolutionary curve $\rho_t$ to determine if $H_t$ generates a geodesic. However, this is not the case, which becomes evident if we go over to the Heisenberg picture. Let $\HHH_t$ be the Hamiltonian in the Heisenberg picture and $P_{j;0}$ be the $j$th eigenspace projector of the initial state $\rho_0$. Then, 
\begin{equation}
	\HHH_t^h = \sum_{j\ne k} P_{j;0} \HHH_t P_{k;0},
\end{equation}
is the horizontal component of $H_t$ in the Heisenberg picture. Furthermore, the geodesic equation \eqref{geodesiceq} reads 
\begin{equation}\label{geodesicHeis}
		\big(\dot \HHH_t^h - i(\HHH_t^h)^2\big)\rho_0 + \rho_0\big(\dot \HHH_t^h + i(\HHH_t^h)^2\big)=0
\end{equation}
in the Heisenberg picture. This equation involves only the Hamiltonian and the initial state.

Equation \eqref{geodesicHeis} suggests how to find a Hamiltonian (in the Schr{\"o}dinger picture) which generates the geodesic that extends from a state $\rho$ with a given velocity $\dot\rho$: First solve \eqref{geodesicHeis} with initial condition $-i[\HHH_t^h,\rho_0]=\dot\rho_0$. Then solve $\dot U_t=-iU_t\HHH_t^h$ with initial condition $U_0=\1$. The parallel transporting Hamiltonian that generates the geodesic which extends from $\rho_0$ with velocity $\dot\rho_0$ is $H_t=i\dot U_t U_t^\dagger$.

\begin{rmk}
If a time-independent Hamiltonian $H$ is horizontal at $\rho_0$, then $H$ is parallel transporting. That is, $H$ is equal to its horizontal component along the entire curve $\rho_t$ generated by $H$ from $\rho_0$. By equation \eqref{geodesiceq}, $\rho_t$ is then a geodesic if and only if $H^2$ commutes with $\rho_0$. Of course, this is a non-generic case. One can also show that if $H_t$ generates a geodesic and if $H_t$ is horizontal for some $t$, then $H_t$ is parallel transporting \cite{AnHe2014}.
\end{rmk}

\begin{ex}
Consider the three nondegenerate qutrits
\begin{align}
	\rho_0 &= p_1\ketbra{1}{1} + p_2\ketbra{2}{2} + p_3\ketbra{3}{3},\\
	\rho_1 &= p_3\ketbra{1}{1} + p_2\ketbra{2}{2} + p_1\ketbra{3}{3},\\
	\rho_2 &= p_2\ketbra{1}{1} + p_3\ketbra{2}{2} + p_1\ketbra{3}{3}, 
\end{align}
where $p_1 > p_2 > p_3$. There is a time-independent parallel transporting Hamiltonian which transforms $\rho_0$ to $\rho_1$ in the time $\tauP$, namely $H= i(\ketbra{3}{1} - \ketbra{1}{3})$. However, there is no time-independent parallel transporting Hamiltonian that transforms $\rho_0$ to $\rho_2$ in the time $\tauP$. For such a Hamiltonian would need to generate a geodesic from $\rho_0$ to $\rho_2$. But no time-independent Hamiltonian whose square commutes with $\rho_0$ can transform $\rho_0$ into $\rho_2$.
\end{ex}

The Grassmann QSL is generally weaker than $\tauP$. However, the Grassmann QSL has the advantage of being explicitly calculable. In cases where the Grassmann QSL is tight it can be used to calculate the geodesic distance with respect to the metric $\gP$.

\begin{ex}
The geodesic distance between two fully distinguishable states is $\pi/2$; cf.\ Example \ref{ex:distinguishable}.
\end{ex}

\begin{ex}
If $\rho_0$ and $\rho_1$ are isospectral and have only two different eigenvalues $p_1$ and $p_2$, then 
\begin{equation}
	\dist_{\gP}(\rho_0,\rho_1)
	=\sqrt{(p_1+p_2)}\distG(P_{1;0},P_{1;1}),
\end{equation}
where $\distG(P_{1;0},P_{1;1})$ is the Grassmann distance between the projectors onto the eigenspaces of $\rho_0$ and $\rho_1$, respectively, corresponding to eigenvalue $p_1$; cf.\ Example \ref{two}.
\end{ex}

\begin{ex}
Let $\rho_0$ and $\rho_1$ be the isospectral, commuting, and nondegenerate mixed states in Example \ref{ex:GrassmanvsUhlmann}, and assume that the permutation $\sigma$ is an involution. Then 
\begin{equation}
	\dist_{\gP}(\rho_0,\rho_1)=\frac{\pi}{2}\sqrt{\sum_{j\ne\sigma(j)} p_j}.
\end{equation}
\end{ex}

\vspace{-12pt}
\section{Summary and outlook}\label{summary}
The Mandelstam-Tamm QSL can be extended in various ways to closed systems in mixed states. In this paper we have derived, analyzed, and compared four such extensions: $\tauG$, $\tauFS$, $\tauU$, and $\tauP$.

Explicit formulas were derived for the first two, making them reasonably easy to apply. By construction, $\tauFS$ is weaker than $\tauG$ but is often easier to calculate. For nondegenerate states, they are the same.

The QSL $\tauU$, originating from Uhlmann's energy dispersion estimate, is well-known and is also explicitly calculable. We analyzed the underlying geometry of Uhlmann's estimate and proved that $\tauU$ is never tight for mixed states of rank greater than half of the system dimension. We also compared $\tauU$ and $\tauG$ in several different cases, and in all of these, $\tauG$ was greater than or equal to $\tauU$. Whether this is the case in general is an open question.

For the fourth QSL, which in a sense is the strongest possible extension of the Mandelstam-Tamm QSL, no explicit formula is known. The fact that Hamiltonians that generate tight evolutions are generally time-varying complicates the derivation of such a formula. If no requirements are imposed on the Hamiltonian, $\tauP$ is tight for every pair of isospectral states; if requirements restrict the set of available Hamiltonians, $\tauP$ no longer needs to be tight. In a forthcoming paper, we will discuss strategies for deriving tight extensions of the Mandelstam-Tamm QSL in such cases.

\onecolumngrid
\appendix

\section{The geodesic distance in Grassmannians}\label{grassmann_distance}
In this appendix we derive the formula \eqref{distanceG} for the geodesic distance between two projectors $P_0$ and $P_1$ in $G(n,\HH)$ equipped with the metric $\gG$ defined in \eqref{Grassmann_metric}.

Let $\hh(\HH)$ be the space of Hermitian operators on $\HH$. The space $\hh(\HH)$ is a parallelizable manifold, and all its tangent spaces can be canonically identified with $\hh(\HH)$ itself. We equip $\hh(\HH)$ with the translation-invariant metric $\gH$ that agrees with half of the Hilbert-Schmidt inner product at the origin, $\gH(H_1,H_2)=\frac{1}{2}\tr(H_1H_2)$.

The Grassmannian $G(n,\HH)$ with the metric $\gG$ is a compact Riemannian submanifold of $\hh(\HH)$. Since the unitary group of $\HH$ acts transitively on $G(n,\HH)$ by left conjugation, every vector in the tangent space of $G(n,\HH)$ at a projector $P$ is of the form $-i[H,P]$ where $H$ is some Hermitian operator on $\HH$. Different $H$s may represent the same tangent vector. We get unique representations if we restrict the set of Hermitian operators to the vector space $\hh(P,\HH)$ consisting of the Hermitian operators on $\HH$ that are represented by block off-diagonal matrices relative to every eigenbasis of $P$:
\begin{equation}
	\hh(P,\HH)=\{H\in\hh(\HH): PHP+(\1-P)H(\1-P)=0\}=\{H\in\hh(\HH): H=HP+PH\}.
\end{equation}
Furthermore, the linear bijection $H \to -i[H,P]$ from $\hh(P,\HH)$ onto the tangent space at $P$ is an isometry:
\begin{equation}\label{isom}
	\gG \big( -i[H_1,P], -i[H_2,P] \big) 
	= \frac{1}{2}\tr\big((-i[H_1,P])(-i[H_2,P])\big)
	= \frac{1}{2}\tr(H_1H_2)
	= \gH (H_1,H_2).
\end{equation}
Notice that this identity holds for $H_1$ and $H_2$ in $\hh(P,\HH)$, but it need not hold for arbitrary Hermitian operators. We say that the Hermitian operators in $\hh(P,\HH)$ are horizontal at $P$.

To determine the geodesic distance between two projectors we first determine what the geodesics in $G(n,\HH)$ look like. A geodesic is a curve whose covariant derivative vanishes identically. The covariant derivative of a curve $P_t$ in $G(n,\HH)$ is the orthogonal projection of the second-order time-derivative $\ddot P_t$, which is the covariant derivative of $P_t$ in $\hh(\HH)$, on the tangent bundle of $G(n,\HH)$. Decompose $\ddot P_t$ as $\ddot P_t = [[\ddot P_t,P_t],P_t] + (\ddot P_t - [[\ddot P_t,P_t],P_t])$. The first component is tangential to $G(n,\HH)$ at $P_t$ since $[\ddot P_t,P_t]$ is skew-Hermitian, and a direct computation shows that the two components are orthogonal as tangent vectors of $\hh(\HH)$. Thus, $[[\ddot P_t,P_t],P_t]$ is the orthogonal projection of $\ddot P_t$ on the tangent space of $G(n,\HH)$ at $P_t$ and, hence, $[[\ddot P_t,P_t],P_t]$ is the covariant derivative of $P_t$. We conclude that $P_t$ is a geodesic if and only if $[[\ddot P_t,P_t],P_t]=0$. The unique solution to this equation that extends from $P$ with velocity $-i[H,P]$, with $H$ being horizontal at $P$, is $P_t=e^{-iHt}P e^{iHt}$. In other words, the geodesics that extend from $P$ are the unitary evolutions of $P$ generated by the Hermitian operators in $\hh(P,\HH)$.

To find an expression for the geodesic distance between $P_0$ and $P_1$ in $G(n,\HH)$ let $P_t$ be a shortest curve from $P_0$ to $P_1$. (Such a curve exists since $G(n,\HH)$ is compact.) Reparameterize $P_t$ if necessary to have a constant speed and to arrive at $P_1$ at $t=1$. 
Then $P_t$ is a geodesic and, consequently, $P_t=e^{-iHt} P_0 e^{iHt}$ for some $H$ in $\hh(P_0,\HH)$. Since $P_t$ is a shortest curve from $P_0$ to $P_1$, the geodesic distance between $P_0$ and $P_1$ is
\begin{equation}\label{kvadratavstand}
	\distG(P_0,P_1)
	= \lengthG[P_t]
	= \sqrt{\frac{1}{2}\tr(H^2)}.
\end{equation}
From the equations characterizing horizontal Hermitian operators at $P_0$ follow that $H^{2k}$ is block diagonal and that $H^{2k+1}$
is block off-diagonal at $P_0$ for every non-negative integer $k$,
\begin{align}
	&H^{2k}=P_0 H^{2k} P_0 + (\1-P_0) H^{2k}(\1-P_0), \label{e1}\\
	&H^{2k+1}=P_0 H^{2k+1}(\1-P_0) + (\1-P_0) H^{2k+1}P_0.\label{e2}
\end{align}
Furthermore,
\begin{align}
	&P_0 H^{2k}P_0 = (P_0 H^k (\1-P_0))(P_0 H^k (\1-P_0))^\dagger,\label{ett}\\
	&(\1-P_0) H^{2k}(\1-P_0) = (P_0 H^k (\1-P_0))^\dagger (P_0 H^k (\1-P_0)).\label{tva}
\end{align}
The two terms on the right-hand side of \eqref{e1} thus have the same nonzero eigenvalues. We conclude that
\begin{equation}\label{fyra}
	\distG^2(P_0,P_1)
	=\frac{1}{2}\tr(H^2)
	=\tr(P_0 H^2P_0).
\end{equation}
According to \eqref{e1} and \eqref{e2}, $\cos H$ is block diagonal, and $\sin H$ is block off-diagonal. Consequently,
\begin{equation}\label{shoji}
	P_0 P_1 P_0=P_0e^{-iH}P_0e^{iH}P_0=(P_0 \cos H P_0)^2. 
\end{equation}
Below we will show that the eigenvalues of $H$ lie between $-\pi/2$ and $\pi/2$. From this follows that $P_0 \cos H P_0$ is positive and hence that
\begin{equation}
	\arccos|P_0 P_1|
	= \arccos (P_0 \cos H P_0)
	= P_0|H|P_0 + \frac{\pi}{2}(\1-P_0).
\end{equation}
Squaring both sides and then taking the trace yields
\begin{equation}\label{elva}
	\tr \arccos^2 |P_0 P_1| 
	= \tr (P_0 H^2 P_0) + \frac{\pi^2}{4}(N-n).
\end{equation}
The formula \eqref{distanceG} follows from \eqref{fyra} and \eqref{elva}:
\begin{equation}
	\distG^2(P_0,P_1)=\tr \arccos^2 |P_0 P_1| - \frac{\pi^2}{4}(N-n).
\end{equation}

It remains to prove that the spectrum of $H$ is contained in the interval $[-\pi/2,\pi/2]$. We will assume that $2n\leq N$. How to treat the case $2n>N$ will be explained afterward. Let $\ket{1},\ket{2},\dots,\ket{N}$ be an orthonormal basis for $\HH$. Write $\supp P_0$ and $\supp (\1-P_0)$ for the support of $P_0$ and $(\1-P_0)$, respectively, and let $F_0$ and $\hat F_0$ be the inclusions of $\supp P_0$ and $\supp (\1-P_0)$ in $\HH$. Define $F_1$ from $\supp P_0$ to $\HH$ and $\hat F_1$ from $\supp (\1-P_0)$ to $\HH$ such that
\begin{alignat}{3}
	&F_1F_1^\dagger = P_1,\qquad && \hat F_1\hat F_1^\dagger = (\1-P_1)
	,\\
	&F_1^\dagger F_1=\1|_{\supp P_0},\qquad && \hat F_1^\dagger \hat F_1=\1|_{\supp (\1-P_0)},
\end{alignat}
where $\1|_{\supp P_0}$ and $\1|_{\supp (\1-P_0)}$ are the identity operators on $\supp P_0$ and $\supp (\1-P_0)$. Equation \eqref{shoji} implies that
\begin{equation}\label{yess}
	F_0^\dagger F_1F_1^\dagger F_0 
	= F_0^\dagger P_0P_1P_0F_0
	= (F_0^\dagger \cos{H} F_0)^2.
\end{equation}
This is the identity obtained from interpreting \eqref{shoji} as an identity between operators on the support of $P_0$. Equation \eqref{yess} tells us that the singular values of $F_0^\dagger F_1$ are $|\cos\eps_1|,|\cos\eps_2|,\dots,|\cos\eps_n|$ where $\eps_1^2,\eps_2^2,\dots,\eps_n^2$ are the eigenvalues of $H^2$ corresponding to the eigenvectors in $\supp P_0$. Let $\xi_j=\arccos |\cos\eps_j|$. Then $0\leq \xi_j\leq |\eps_j|$, and $\xi_j=|\eps_j|$ only if $|\eps_j|\leq\pi/2$. Next we will show that the distance from $P_0$ to $P_1$ does not exceed $(\xi_1^2+\xi_2^2+\dots+\xi_n^2)^{1/2}$ by constructing a curve $R_t$ connecting $P_0$ and $P_1$ that has exactly this length. It then follows from the assumption that $P_t$ has a minimum length that the spectrum of $H$ is contained in $[-\pi/2,\pi/2]$.  

Let $\ket{1},\ket{2},\dots,\ket{n}$ and $\ket{n+1},\ket{n+2},\dots,\ket{N}$ be orthonormal bases in $\supp P_0$  and $\supp (\1-P_0)$, respectively. Define $W$ on $\HH$ by the assumptions that $W=F_1$ on $\supp P_0$ and $W=\hat F_1$ on $\supp (\1-P_0)$. Then $W$ is unitary and $F_0^\dagger W F_0=F_0^\dagger F_1$. According to the CS-decomposition theorem, see \cite{Bh1997}, there exist unitary operators $U$ and $V$ on $\supp P_0$, and $\hat U$ and $\hat V$ on $\supp (\1-P_0)$, such that $Q=(U\oplus \hat U)^\dagger W (V\oplus \hat V)$ satisfies 
\begin{alignat}{2}
	&\bra{k} Q \ket{l}=\delta_{kl} \cos(\xi_k) \qquad && \text{if} \quad 1\leq k,l\leq n,\\
	&\bra{k} Q \ket{n+l}=-\delta_{kl} \sin(\xi_k) \qquad && \text{if} \quad 1\leq k,l\leq n,\\
	&\bra{n+k} Q \ket{l}=\delta_{kl} \sin(\xi_k) \qquad && \text{if} \quad 1\leq k,l\leq n,\\
	&\bra{n+k} Q \ket{n+l}=\delta_{kl} \cos(\xi_k) \qquad && \text{if} \quad 1\leq k,l\leq n,\\
	&\bra{n+k} Q \ket{n+l}=\delta_{kl} \qquad && \text{if} \quad k>n\;\text{or}\; l>n.
\end{alignat}
Define $Q_t$ by
\begin{alignat}{2}
	&\bra{k} Q_t \ket{l}=\delta_{kl} \cos(t\xi_k)\hspace{7pt} && \text{if}\hspace{5pt} 1\leq k,l\leq n,\\
	&\bra{k} Q_t \ket{n+l}=-\delta_{kl} \sin(t\xi_k)\hspace{7pt} && \text{if}\hspace{5pt} 1\leq k,l\leq n,\\
	&\bra{n+k} Q_t \ket{l}=\delta_{kl} \sin(t\xi_k)\hspace{7pt} && \text{if}\hspace{5pt} 1\leq k,l\leq n,\\
	&\bra{n+k} Q_t \ket{n+l}=\delta_{kl} \cos(t\xi_k)\hspace{7pt} && \text{if}\hspace{5pt} 1\leq k,l\leq n,\\
	&\bra{n+k} Q_t \ket{n+l}=\delta_{kl} \hspace{7pt} && \text{if}\hspace{5pt} k>n\;\text{or}\; l>n.
\end{alignat}
and define $W_t=(U\oplus \hat U)Q_t(V\oplus \hat V)^\dagger$. Also, for $0\leq t\leq 1$, define $R_t=W_tP_0W_t^\dagger$. The operator $W_t$ is unitary and, hence, $R_t$ is a curve in $G(n,\HH)$. The curve starts at $R_0 = (UV^\dagger \oplus \hat U\hat V^\dagger) P_0 (VU^\dagger \oplus \hat V\hat U^\dagger) = P_0$ and finishes at $R_1 = W F_0 F_0^\dagger W^\dagger = F_1F_1^\dagger = P_1$. Furthermore, the speed of $R_t$ squared is $\|\dot R_t\|_{\gG}^2=\tr(\dot Q_t^\dagger \dot Q_tP_0)=\xi_1^2+\xi_2^2+\dots+\xi_d^2$. We conclude that the length of $R_t$ is $\sqrt{\xi_1^2+\xi_2^2+\dots+\xi_d^2}$. Observe that $R_t$ is shorter than $P_t$ unless $\xi_j=|\eps_j|$. This proves that the spectrum of $H$ is contained in $[-\pi/2,\pi/2]$.

The application of the CS-decomposition theorem required that $2n\leq N$. If $2n > N$, the projectors $P_0$ and $P_1$ have a common invariant subspace $\HH_0$ of dimension at least $2n-N$. Decompose $\HH$ as $\HH_0\oplus\HH_0^\bot$, where $\HH_0^\bot$ is the orthogonal complement of $\HH_0$. Then, if we restrict $P_0$ and $P_1$ to $\HH_0^\bot$ and regard the restrictions as operators on $\HH_0^\bot$, the CS-decomposition argument can be applied to produce a curve of projectors $R_t$ on $\HH_0^\bot$ which can be extended to all of $\HH$ by assuming that it acts trivially on $\HH_0$. The curve $R_t$ is strictly shorter than $P_t$ if the spectrum of $H$ is not contained in $[-\pi/2,\pi/2]$.

\section{The Pl{\"u}cker embedding is an isometry}\label{plucker_is_isometry}
In this appendix, we show that the Pl{\"u}cker embedding is an isometry. Specifically, we show that if $P_t$ is a curve in $G(n,\HH)$ and $F_t$ is a curve of frames for the $P_t$s, then $\|\dot P_t\|^2_{\gG}= \braket{\dot F_t}{\dot F_t} - |\braket{ F_t}{\dot F_t}|^2$; compare with the formula \eqref{snok}. The notation $\ket{\dot F_t}$ for the velocity of $\ket{F_t}$ is somewhat ambiguous. To make precise what we mean write $F_t=(\ket{1_t}\,\ket{2_t}\cdots\ket{n_t})$. Then $\ket{\dot F_t}=\sum_{k=1}^n \ket{1_t}\wedge \cdots \wedge \ket{\dot k_t} \wedge \cdots \wedge\ket{n_t}$.

Let $H_t$ be a Hamiltonian generating $P_t$, and without loss of generality assume that $H_t$ is horizontal at $P_t$ for every $t$; see Appendix \ref{grassmann_distance}. (Otherwise, replace $H_t$ with $H_tP_t+P_tH_t$.) Take $F_t$ to be such that $\dot F_t=-iH_tF_t$. The benefit of choosing the Hamiltonian horizontal is that each $\ket{\dot k_t}$ is then perpendicular to all the vectors in the frame $F_t$. Denote the row matrix obtained by removing the $k$th vector from $F_t$ by $F_t^k$ and the row matrix obtained by removing both the $k$th and the $l$th vector from $F_t$ by $F_t^{kl}$. Since $\ket{\dot k_t}$ is perpendicular to all the vectors in $F_t$ we have that $\big\langle \ket{\dot k_t} \wedge \ket{F^k_t} \big| \ket{F_t} \big\rangle=0$. Also, if $k\ne l$, then $\big\langle \ket{\dot k_t} \wedge \ket{l_t} \wedge  \ket{F^{kl}_t} \big| \ket{\dot l_t} \wedge \ket{k_t} \wedge\ket{F^{kl}_t}\big\rangle = 0$. This is so because, for example, the second row of the matrix of which we take the determinant when calculating this inner product contains only zeroes. It follows that
\begin{equation}
	\braket{\dot F_t}{F_t}=\sum_{k=1}^n  (-1)^{k-1}
	\big\langle \ket{\dot k_t} \wedge \ket{F^k_t} \big| \ket{F_t} \big\rangle=0
\end{equation}
and that
\begin{equation}
\begin{split}
	\braket{\dot F_t}{\dot F_t}
	&=\sum_{k=1}^n (-1)^{2(k-1)} \big\langle \ket{\dot k_t} \wedge \ket{F^k_t} \big| \ket{\dot k_t} \wedge \ket{F^k_t} \big\rangle
	-\sum_{k=1}^n \sum_{l\ne k} (-1)^{k+l}
	\big\langle \ket{\dot k_t} \wedge \ket{l_t} \wedge  \ket{F^{kl}_t} \big| \ket{\dot l_t} \wedge \ket{k_t} \wedge\ket{F^l_t}\big\rangle \\
	&=\sum_{k=1}^n \big\langle \ket{\dot k_t} \wedge \ket{F^k_t} \big| \ket{\dot k_t} \wedge \ket{F^k_t} \big\rangle\\	
	&=\sum_{k=1}^n \braket{\dot k_t}{\dot k_t}. 
\end{split}
\end{equation}
Two times the square of the Grassmann speed of $P_t$ is $\tr(H^2_t)$, and arguments identical to those that led to equation \eqref{fyra} show that $\tr(H^2_t) = 2\tr(P_tH_t^2P_t)$. We conclude that 
\begin{equation}
	\braket{\dot F_t}{\dot F_t} - |\braket{ F_t}{\dot F_t}|^2
	= \sum_{k=1}^n\braket{\dot k_t}{\dot k_t}
	= \tr(P_tH_t^2P_t)
	= \frac{1}{2}\tr(H_t^2)
	= \|\dot P_t\|_{\gG}^2.
\end{equation}

\section{The Bures angle}\label{Uhlmanns_theorem}
Here we show that the distance between the fibers over $\rho_0$ and $\rho_1$ in the Uhlmann bundle is $\arccos\tr|\sqrt{\rho_0}\sqrt{\rho_1}|$. According to equation \eqref{call}, this follows from the observation that for any two amplitudes $W_0$ and $W_1$ for $\rho_0$ and $\rho_1$,
\begin{equation}
	\max_{U\in \U(n)}\Re\tr(W_0^\dagger W_1 U)=\tr|W_0^\dagger W_1|=\tr|\sqrt{\rho_0}\sqrt{\rho_1}|.
\end{equation}
Notice that $\sqrt{\rho_0}$ and $\sqrt{\rho_1}$ are not amplitudes for $\rho_0$ and $\rho_1$ unless $\rho_0$ and $\rho_1$ are faithful.

The first identity follows from a well-known calculation involving the polar representation and the Cauchy-Schwarz inequality; see \cite{NiCh2010}. Write $W_0^\dagger W_1=|W_0^\dagger W_1|V$ where $V$ belongs to $\U(n)$. Then, for every $U$ in $\U(n)$,
\begin{equation}
\begin{split}
	\Re\tr(W_0^\dagger W_1 U)
	&\leq |\tr(W_0^\dagger W_1 U)| \\
	&\leq |\tr(|W_0^\dagger W_1|^{\frac{1}{2}} |W_0^\dagger W_1|^{\frac{1}{2}} V U)| \\
	&\leq \sqrt{\tr|W_0^\dagger W_1|}\sqrt{\tr(U^\dagger V^\dagger |W_0^\dagger W_1| VU)} \\
	&=\tr|W_0^\dagger W_1|.
\end{split}
\end{equation}
Equality is obtained if we choose $U=V$.

The next step is to prove that $|W_0^\dagger W_1|$ and $|\sqrt{\rho_0}\sqrt{\rho_1}|$ have the same nonzero eigenvalues, or equivalently that $W_0^\dagger \rho_1 W_0$ and $\sqrt{\rho_0}\rho_1\sqrt{\rho_0}$ have the same nonzero eigenvalues. For this, let $\ket{u_1},\ket{u_2},\dots,\ket{u_n}$ be pairwise orthogonal and normalized eigenvectors of $\rho_0$ with nonzero eigenvalues, and let $\ket{k}=\bra{u_k}\rho_0\ket{u_k}^{-1/2} W_0^\dagger\ket{u_k}$. Then $\ket{1},\ket{2},\dots,\ket{n}$ is an orthonormal basis in $\HH^n$. Extend $W_0^\dagger\rho_1W_0$ to an operator on $\HH$ by assuming that $W_0^\dagger\rho_1W_0$ acts as the zero operator on the orthogonal complement of $\HH^n$. Also, let $U$ be a unitary operator on $\HH$ such that $U\ket{u_k}=\ket{k}$. Then $U^\dagger W_0^\dagger \rho_1 W_0 U$ and $\sqrt{\rho_0}\rho_1\sqrt{\rho_0}$ act trivially on the orthogonal complement of the support of $\rho_0$. Furthermore, on the support of $\rho_0$, which is spanned by the $\ket{u_k}$s,
\begin{equation}
	\bra{u_k} U^\dagger W_0^\dagger \rho_1 W_0 U \ket{u_l}
	=\bra{k} W_0^\dagger \rho_1 W_0 \ket{l}
	=\frac{\bra{u_k} \rho_0 \rho_1 \rho_0 \ket{u_l}}{\sqrt{\bra{u_k}\rho_0\ket{u_k}\bra{u_l}\rho_0\ket{u_l}}} 
	=\bra{u_k}\sqrt{\rho_0} \rho_1 \sqrt{\rho_0}\ket{u_l}. 
\end{equation}
This proves that the extension of $W_0^\dagger \rho_1 W_0$ is isospectral to $\sqrt{\rho_0} \rho_1 \sqrt{\rho_0}$ and hence that $|W_0^\dagger W_1|$ and $|\sqrt{\rho_0}\sqrt{\rho_1}|$ have the same nonzero eigenvalues.

\section{A legitimate modification of the Hamiltonian}\label{horizontal_modification}
Assume that $\rho_t$, generated by $H_t$, is locally length minimizing with respect to the Bures metric and has a speed that equals the energy uncertainty. Also, assume that $\rho_t$ is mixed, that is, has a rank which is greater than $1$. Then $H_t$ can be reparameterized so that it generates a reparameterized version of $\rho_t$ with a constant speed that equals the uncertainty of the reparameterized Hamiltonian. Furthermore, the reparameterized Hamiltonian can be modified to generate a horizontal lift of the reparameterization of $\rho_t$ in the Uhlmann amplitude bundle: Since $\rho_t$ is mixed and has a speed equal to the energy uncertainty, $\rho_t$ is never stationary. From this follows that $\rho_t$ can be reparameterized according to arclength; see \cite{Sa1996}. Let $t=t(s)$ be the arclength parameterization and define $\varrho_s=\rho_{t(s)}$ and $h_s=t'(s)H_{t(s)}$. Here $'$ means differentiation with respect to $s$. Then $\varrho_s'=-i[h_s,\varrho_s]$ and $\|\varrho_s'\|_{\gB}=1$. Since $\rho_t$ is locally length minimizing, so is $\varrho_s$. But then, as $\varrho_s$ has a constant speed, $\varrho_s$ is a geodesic. Furthermore,
\begin{equation}
	\|\varrho_s'\|^2_{\gB}
	= t'(s)^2\|\dot\rho_{t(s)}\|^2_{\gB} 
	= t'(s)^2\Delta(H_{t(s)},\rho_{t(s)}) 
	= \Delta^2(t'(s)^2 H_{t(s)},\rho_{t(s)}) 
	= \Delta^2(h_s,\varrho_s).
\end{equation}
Finally, according to equation \eqref{just}, a lift $W_s$ of $\varrho_s$ satisfying $W_s'=-ih_sW_s$ must be such that $\AU(W_s')=-i\alpha_s\1$ for some real-valued function $\alpha_s$. Replace $h_s$ with $h_s-\alpha_s\1$. Then $W_s$, satisfying $W_s'=-i(h_s-\alpha_s\1)W_s$, is a horizontal lift of $\varrho_s$. Since $\varrho_s$ is a geodesic, $W_s$ is also a geodesic.
\end{document}